\documentclass[fleqn,usenatbib]{mnras}
\usepackage{newtxtext,newtxmath}
\usepackage[T1]{fontenc}

\DeclareRobustCommand{\VAN}[3]{#2}
\let\VANthebibliography\thebibliography
\def\thebibliography{\DeclareRobustCommand{\VAN}[3]{##3}\VANthebibliography}

\usepackage{graphicx}	
\usepackage{amsmath}	
\usepackage{threeparttable} 
\usepackage{tabularx} 
\usepackage{color} 
\usepackage{float} 

\newcommand{\xflux}{erg cm$^{-2}$ s$^{-1}$\ }   
\newcommand{\xmm}{\textit{XMM-Newton}\ }        
\newcommand{\Rearth}{R_\text{E}}
\newcommand{\Mearth}{M_\text{E}}

\newcommand{\code}[1]{\texttt{#1}}

\newcommand{\narrowhline}[1]{\hline & \\[-#1ex] \ }

\title[The shared evaporation history of three sub-Neptunes]{The shared evaporation history of three sub-Neptunes spanning the radius--period valley of a Hyades star}

\author[J. Fernández Fernández et al.]{
Jorge Fernández Fernández,$^{1,2}$\thanks{E-mail: jorge.fernandez-fernandez@warwick.ac.uk}
Peter J. Wheatley,$^{1,2}$\thanks{E-mail: p.j.wheatley@warwick.ac.uk}
George W. King$^{3,1,2}$\thanks{E-mail: kinggw@umich.edu}
\\
$^{1}$Department of Physics, University of Warwick, Gibbet Hill Road, Coventry CV4 7AL, UK\\
$^{2}$Centre for Exoplanets and Habitability, University of Warwick, Gibbet Hill Road, Coventry CV4 7AL, UK\\
$^{3}$Department of Astronomy, University of Michigan, Ann Arbor, MI 48109, USA
}

\date{Accepted XXX. Received YYY; in original form ZZZ}

\pubyear{2023}

\begin{document}
\label{firstpage}
\pagerange{\pageref{firstpage}--\pageref{lastpage}}
\maketitle

\begin{abstract}
We model the evaporation histories of the three planets around K2-136,
a K-dwarf in the Hyades open cluster with an age of 700\,Myr.  The star hosts three transiting planets, with radii of 1.0, 3.0 and 1.5 Earth radii, where the middle planet lies above the radius--period valley and the inner and outer planets are below. We use an \xmm\ observation to measure the XUV radiation environment of the planets, finding that the X-ray activity of K2-136 is lower than predicted by models but typical of similar Hyades members. We estimate the internal structure of each planet, and model their evaporation histories using a range of structure and atmospheric escape formulations. While the precise X-ray irradiation history of the system may be uncertain, we exploit the fact that the three planets must have shared the same history. We find that the Earth-sized K2-136b is most likely rocky, with any primordial gaseous envelope being lost within a few Myr. The sub-Neptune, K2-136c, has an envelope contributing 1--1.7\% of its mass that is stable against evaporation thanks to the high mass of its rocky core, whilst the super-Earth, K2-136d, must have a mass at the upper end of the allowed range in order to retain any of its envelope. Our results are consistent with all three planets beginning as sub-Neptunes that have since been sculpted by atmospheric evaporation to their current states, stripping the envelope from planet b and removing most from planet d whilst preserving planet c above the radius-period valley.
\end{abstract}



\section{Introduction}

The \textit{Kepler} mission uncovered a large population of close-in exoplanets between Earth and Neptune in size \citep[1--4\,$\Rearth$;][]{borucki11}, raising questions as to their formation and evolution. \citet{fulton-2017} found that the radii of these planets form a bimodal distribution with a valley at about 1.8 $\Rearth$, and \cite{van-eylen-2018} showed that this valley moves to smaller radii for longer orbital periods. Furthermore, mass measurements from radial velocity follow up \citep[e.g.][]{rv-gj-3998-affer-2016, rv-toi-561-lacedelli-2022} and transit timing variations \citep[TTVs; e.g.][]{kepler-19-ttv-ballard-2011, kepler-ttv-steffen-2013, trappist-one} have shown that planets with radii below $\sim$2 $\Rearth$ tend to have densities consistent with an Earth-like (rocky) composition, whereas those planets above the gap have lower densities consistent with gaseous envelopes. 

H/He gaseous envelopes on low-mass planets can double the radius of a planet while comprising less than one percent of its mass. Moreover, these envelopes can undergo escape from Neptunes and sub-Neptunes, which has been observed as planetary tails of escaping gas in the Lyman-alpha line of hydrogen 
\citep{lyman-atm-ehrenreich-2015,DosSantos20:sub-neptune-escape,lyman-atm-zhang-2022}, 
and more recently using helium absorption 
\citep{Allart18:neptune-helium-escape, Zhang23:neptune-helium-escape}.
The radius-period valley is thus thought to be consistent with a picture in which atmospheric escape plays an important role in sculpting the population of close-in small exoplanets, with some planets having their envelopes completely stripped down to their core, as predicted by \citet{lopez-fortney-2013}, \citet{owen-wu-2013} and \citet{jin-2014}.

In these models, the atmospheric escape is driven by stellar X-ray and extreme-ultraviolet (EUV) radiation (together XUV), although other mechanisms have also been proposed. For instance, \citet{wyatt20} argued that impacts by planetesimals could also have shaped the radius valley, and \citet{ginzburg-2016} proposed core-powered mass loss, where the planet's internal cooling luminosity drives evaporation, operating on Gyr timescales. Indeed, \citet{ginzburg18} and \citet{gupta19} showed that core-powered mass loss can also reproduce the radius-period valley.

XUV emission from late-type stars is a result of coronal heating driven by a magnetic dynamo mechanism, which is powered by stellar rotation. Indeed, the X-ray activity of a star is tied to its rotation period, with faster rotators being more luminous in the X-ray regime up to a saturation limit at $L_\text{X} / L_\text{bol}\sim10^{-3}$, where the relation flattens \citep[e.g.][]{pallavicini-1981, pizzolato-2003, wright-2011}. The incoming XUV radiation heats the upper layers of exoplanet atmospheres, driving a hydrodynamic wind that overflows the Roche lobe and escapes. 

Stellar X-ray luminosity weakens with time as angular momentum loss through the stellar wind causes stars to spin down \citep{jackson-2012, tu-2015, johnstone-2021}, and by the age of 1 Gyr hard X-ray emission has been reduced by a factor of 100. The softer EUV emission is thought to decline more slowly, and can continue to be significant for several Gyr \citep{king-2021}.

Observations, however, only provide a snapshot of a planet's evolutionary history. One can attempt to model a planet's current physical properties but its history can be hard to constrain. A straightforward solution to break the degeneracy is to study multiplanetary systems, where all the planets in the system share the same the XUV irradiation history. Their current states can then be used to constrain each other's pasts, especially if the planets are found on both sides of the radius valley \citep{campos-estrada-20}.

In this paper we study the evaporation history of three planets around the star K2-136 (EPIC 247589423, LP 358-348), a K5V dwarf at a distance of 59\,pc. The star, whose parameters are shown in Table \ref{tab:star_params}, is a member of the Hyades cluster, which makes it about 700 Myr in age \citep{hyades-age-brandt-2015, hyades-age-martin-2018}. The three planets (K2-136\,b, c, and d) have radii 1.0, 3.0, and 1.5 $\Rearth$, respectively, as reported by \citet{ciardi-2018}, \citet{livingstone-2018}, and \citet{mann-2017}, and their orbital and physical properties are shown in Table \ref{tab:planets}. Given that the planets straddle the radius valley, this system provides an excellent opportunity to study the evaporation history of exoplanets in a shared X-ray environment.

The paper is structured as follows: Sect.\,\ref{sec:planets} introduces the planets in the system and their physical properties, Sect.\,\ref{sec:xray} concerns our X-ray observations of K2-136, in Sects.\,\ref{sec:sim}\,\&\,\ref{sec:results} we model the evaporation histories of the three planets, and in Sects.\,\ref{sec:discussion}\,\&\,\ref{sec:conclusion} we discuss and summarise our results.

\begin{table}
    \centering
    \caption{Stellar parameters for K2-136.}
    \begin{threeparttable}
    \begin{tabular}{llcc}
        \hline\hline
        Parameter & Units & Value & Reference \\
        \narrowhline{5} \\
        \multicolumn{4}{c}{Spatial} \\
        RA  (J2000)         & hh:mm:ss      & $04:29:38.99$     & Gaia DR3\tnote{b} \\
        Dec (J2000)         & dd:mm:ss      & $+22:52:57.78$    & Gaia DR3\tnote{b} \\
        $\mu_{\text{RA}}$   & mas yr$^{-1}$ & $82.778\pm0.021$  & Gaia DR3\tnote{b} \\
        $\mu_{\text{Dec}}$  & mas yr$^{-1}$ & $-35.541\pm0.015$ & Gaia DR3\tnote{b} \\
        Parallax            & mas           & $16.982\pm0.019$  & Gaia DR3\tnote{b} \\
        Distance            & pc            & $58.89\pm0.065$   & Gaia DR3\tnote{b} \\
        \narrowhline{5} \\
        \multicolumn{4}{c}{Photometric} \\
        B   & mag       & $12.48\pm0.01$          & UCAC4\tnote{a} \\
        V   & mag       & $11.20\pm0.01$          & UCAC4\tnote{a} \\
        G   & mag       & $10.8537\pm0.0005$      & Gaia DR3\tnote{b} \\
        BP  & mag       & $11.5398\pm0.001$       & Gaia DR3\tnote{b} \\
        RP  & mag       & $10.0527\pm0.001$       & Gaia DR3\tnote{b} \\
        J   & mag       & $9.096\pm0.022$         & 2MASS\tnote{d} \\
        H   & mag       & $8.496\pm0.020$         & 2MASS\tnote{d} \\
        Ks  & mag       & $8.368\pm0.019$         & 2MASS\tnote{d} \\
        \narrowhline{5} \\
        \multicolumn{4}{c}{Physical} \\
        Spectral type              & ---         & $K5V$      & L18\tnote{e} \\
        T$_{\text{eff}}$ & K           & $4359 \pm 50$        & L18\tnote{e} \\
        M$_*$            & $M_{\odot}$ & $0.686 \pm 0.028$    & L18\tnote{e} \\
        R$_*$            & $R_{\odot}$ & $0.723 \pm 0.072$    & L18\tnote{e} \\
        L$_*$            & $L_{\odot}$ & $0.171 \pm 0.036$    & L18\tnote{e} \\
        {[Fe/H]}         & dex         & $0.17 \pm 0.12$      & L18\tnote{e} \\
        P$_{\text{rot}}$ & days        & $13.6^{+2.2}_{-1.5}$ & L18\tnote{e} \\
        P$_{\text{rot}}$ & days        & $13.8\pm1.0$         & \citet{ciardi-2018} \\
        P$_{\text{rot}}$ & days        & $15.0\pm1.0$         & \citet{mann-2017} \\
        Age              & Myr         & 625$\pm$50           & \citet{hyades-age-perryman-1998} \\
        Age              & Myr         & 750$\pm$100          & \citet{hyades-age-brandt-2015} \\
        Age              & Myr         & 650$\pm$70           & \citet{hyades-age-martin-2018} \\
        \hline
    \end{tabular}
    \begin{tablenotes}
        \item[a] UCAC4: \citet{simbad-ucac4}
        \item[b] Gaia DR3: \citet{gaia-dr3}
        \item[d] 2MASS: \citet{2mass}
        \item[e] L18: \citet{livingstone-2018}
    \end{tablenotes}
    
    \label{tab:star_params}
     \end{threeparttable}
\end{table}

\begin{table*}
    \centering
    \caption{Properties of the transiting exoplanets orbiting K2-136. Radius estimates are divided in three columns, following the source studies, from left to right:
    \citet[L18]{livingstone-2018}, \citet[M18]{mann-2017}, and \citet[C18]{ciardi-2018}. The radii quoted on the discovery papers are shown on the three leftmost columns. The radii we calculated using $R_\text{p}/R_*$ values from these papers together with the Gaia DR3 estimated stellar radius are shown on the three middle columns.
    The last three columns are orbital periods and separations from \citet{mann-2017} and planet masses from 
    \citet[][M23]{Mayo23:k2-136-masses}.}
    \addtolength{\tabcolsep}{-1pt} 
    \renewcommand{\arraystretch}{1.2} 
    \begin{threeparttable}
    \begin{tabular}{c c c c c c c c c c}
        \hline \hline
        Planet & \multicolumn{3}{|c|}{Radius ($\Rearth$)} & \multicolumn{3}{c|}{Radius ($\Rearth$) updated using Gaia DR3} & Orbital Period (days) 
        & Semi-major axis (AU) & Mass ($\Mearth$) \\
        \hline
            & L18 & M18 & C18 & L18 & M18 & C18 
            & M18 & M18 & M23 \\
        
        b   & 1.05$\pm$0.16             
            & 0.99$^{+0.06}_{-0.04}$    
            &                           
            & $1.00\pm0.05$             
            & $1.02\pm0.04$             
            &                           
            & 7.97520 $\pm$ 0.00079     
            & $0.071^{+0.008}_{-0.004}$ 
            & <2.9\tnote{a} \hspace{0.1cm} <4.3\tnote{b} \\ 
        
        c   & 3.14$\pm$0.36             
            & 2.91$^{+0.11}_{-0.10}$    
            & 3.03$^{+0.53}_{-0.47}$    
            & $2.98\pm0.05$             
            & $3.00\pm0.07$             
            & $2.88\pm0.01$             
            & 17.30713 $\pm$ 0.00027    
            & $0.120^{+0.022}_{-0.006}$ 
            & 18.1$\pm$1.9 \\           
        
        d   & 1.55$^{+0.21}_{-0.24}$    
            & 1.45$^{+0.11}_{-0.08}$    
            &                           
            & $1.47\pm0.07$             
            & $1.50\pm0.07$             
            &                           
            & 25.5750 $\pm$ 0.0024      
            & $0.156^{+0.017}_{-0.015}$ 
            & <1.3\tnote{a} \hspace{0.1cm} <3.0\tnote{b} \\ 
        \hline
    \end{tabular}
    \footnotesize
    \begin{tablenotes}
    \item[a] 1$\sigma$ upper limit
    \item[b] 2$\sigma$ upper limit
    \end{tablenotes}
    \label{tab:planets}
    \end{threeparttable}
\end{table*}

\section{Planetary system}
\label{sec:planets}

The planetary radii, orbital periods, and mass measurements of the three planets around K2-136 are shown in Table \ref{tab:planets}. The radius measurements from the different studies all agree within $1\sigma$, and the orbital periods all agree within minutes. 

The precision of these original radius measurements were limited by the precision of the stellar radius, which in turn was limited by knowledge of the distance to K2-136. Since the publication of the discovery papers, we now have improved precision on the stellar radius from Gaia, and
we recalculated the planet radii using the $R_\text{p}/R_*$ values from the discovery papers together with the stellar radius of $0.6854\pm0.0010\,R_\odot$ from Gaia DR3 \citep{gaia-dr3}. This resulted in reduced uncertainties as well as closer consistency between the radii from \citet{mann-2017} and \citet{livingstone-2018}
(see Table\,\ref{tab:planets}).

The mass measurements in Table\,\ref{tab:planets} are from HARPS-N observations. \citet{mayo-2021:abstract} published an abstract summarising these HARPS-N measurements for K2-136, where they report a mass of $15.9\pm2.4\,M_\text{E}$ for K2-136\,c. However, during the preparation of this paper we became aware of revised HARPS-N masses and limits for all three planets
by \citet{Mayo23:k2-136-masses},
which we show in Table \ref{tab:planets}.

Despite the relatively high mass of K2-136\,c of $18.1\pm1.9\,\Mearth$, the planet still has a relatively low density of $\sim3.8\,\rm g\,cm^{-3}$ (using $R_\text{p} = 3.0\,\Rearth$) strongly suggesting the presence of a gaseous envelope. Since envelopes tend to make up a small percentage of the mass of small planets, we can assume that the total mass is approximately equal to the core mass. We then use the mass-radius relations by \citet{otegi-2020} for rocky planets
to estimate a core radius for K2-136\,c of $2.38\pm0.11\,\Rearth$, making its envelope $0.58\pm0.18\,\Rearth$ in thickness.

K2-136\,d, with a radius of $1.48\pm0.11\,\Rearth$, has mass upper limits of $1.3\,\Mearth$ (1$\sigma$) and $3.0\,\Mearth$ (2$\sigma$). Assuming no envelope, the mass-radius relation by \citet{otegi-2020} suggests a planet mass of $3.49\pm0.94\,\Mearth$, which is consistent with the 2$\sigma$ mass upper limit. The 1$\sigma$ limit, however, results in a density of $2.1\,\rm g\,cm^{-3}$, which suggests the presence of an envelope. For this case, using the analysis above, we estimate a core radius of $1.10\,\Rearth$ and thus an envelope size of $0.4\,\Rearth$. We therefore choose to explore the evaporation past of K2-136\,d using a range of core masses between these two scenarios.

Finally, the small size of K2-136\,b, placing it well below the radius valley, and its close proximity to the host star indicates that it lacks an envelope. We adopt a mass of $0.95\,\Mearth$, estimated with the mass-radius relations by \citet{otegi-2020}, which is well within the 1$\sigma$ and 2$\sigma$ upper limits of $2.9$ and $4.3\,\Mearth$, respectively.

\begin{figure}
	\includegraphics[width=\columnwidth]{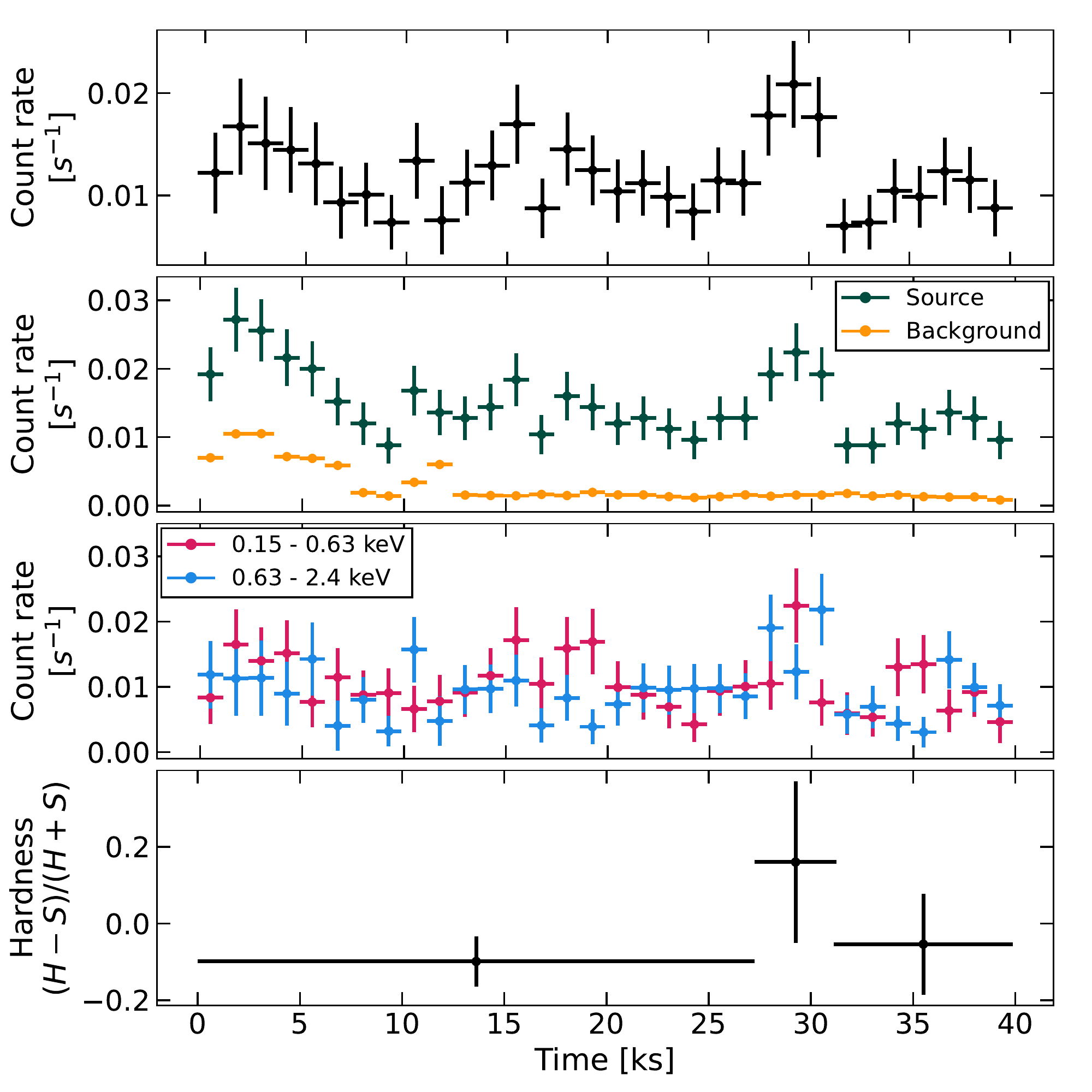}
    \caption{\textbf{Top panel}: background-subtracted lightcurve of K2-136. \textbf{Middle-top panel}: X-ray lightcurve of K2-136, with background count rates in yellow and source count rates in dark green. \textbf{Middle-bottom panel}: background-corrected count rate for soft (0.15-0.63 keV) and hard (0.63-2.4 keV) X-ray counts in red and blue, respectively. \textbf{Bottom panel}: hardness $(H-S)/(H+S)$ between the soft ($S$) and hard ($H$) lightcurves.}
    \label{fig:lightcurve}
\end{figure}

\section{X-ray observations \& analysis}\label{sec:xray}

We observed K2-136 with the \xmm telescope  on 2018 November 9, from 11:14:01 to 23:10:41
(observation ID 08248 50201), 
with a total exposure time of 41 ks, using the European Photon Imaging Camera (EPIC). A preliminary inspection of the data revealed some contamination by particle background flares in the early part of the observation, which are common in \xmm observations \citep{walsh-2014:xmm}. 
However, the signal to noise ratio of the spectrum was not improved by rejecting these events and so we chose not to filter out these time intervals. 

The data were reduced using the Science Data Analysis System (SAS)\footnote{\url{https://www.cosmos.esa.int/web/xmm-newton/sas-threads}}, a collection of digital procedures and libraries designed to analyse the \xmm telescope outputs.
Since our target is a soft X-ray source we used the data from the EPIC-pn detector, which has an operative range of 0.15 to 12 keV and is the most sensitive to soft X-rays \citep{epic-pn}.
We extracted the source counts using a circular region 15\,arcsec in radius centred on the proper-motion corrected coordinates of the star, and we estimated the background using a circular aperture of radius 148\,arcsec on a nearby source-free region.
A search of Gaia DR3 \citep{gaia-dr3} reveals no other sources within a 30 arcsecond radius.

\subsection{Light curve}

The background-subtracted lightcurve is plotted in Fig.\,\ref{fig:lightcurve} (topmost panel). We found an average background-subtracted count rate of $0.012\rm\,$\,s$^{-1}$ for counts in the energy band 0.15 -- 2.4\,keV.
The first 12\,ks of the observation has elevated background counts. However, there is no increase in the background count rate at the time of the source brightening 30\,ks into the observation, which is likely caused by an X-ray flare from the star. 

We calculated the energy at which half the photon counts are found at lower (softer) energies and the other half at higher (harder) energies, which we found to be $0.63$\,keV.
We hence plotted the soft (0.15--0.63\,keV) and hard (0.63-2.40\,keV) X-ray lightcurves separately on Fig.\,\ref{fig:lightcurve}, as well as the hardness ratio, defined as $(H-S)/(H+S)$, which is binned to isolate the potential flare at around 30\,ks. We found marginal evidence that the source is harder at the time of this event, which would be expected for a stellar flare.

\subsection{Spectral analysis}
\label{sec:xray-spectrum}

\begin{figure}
    \includegraphics[width=\columnwidth]{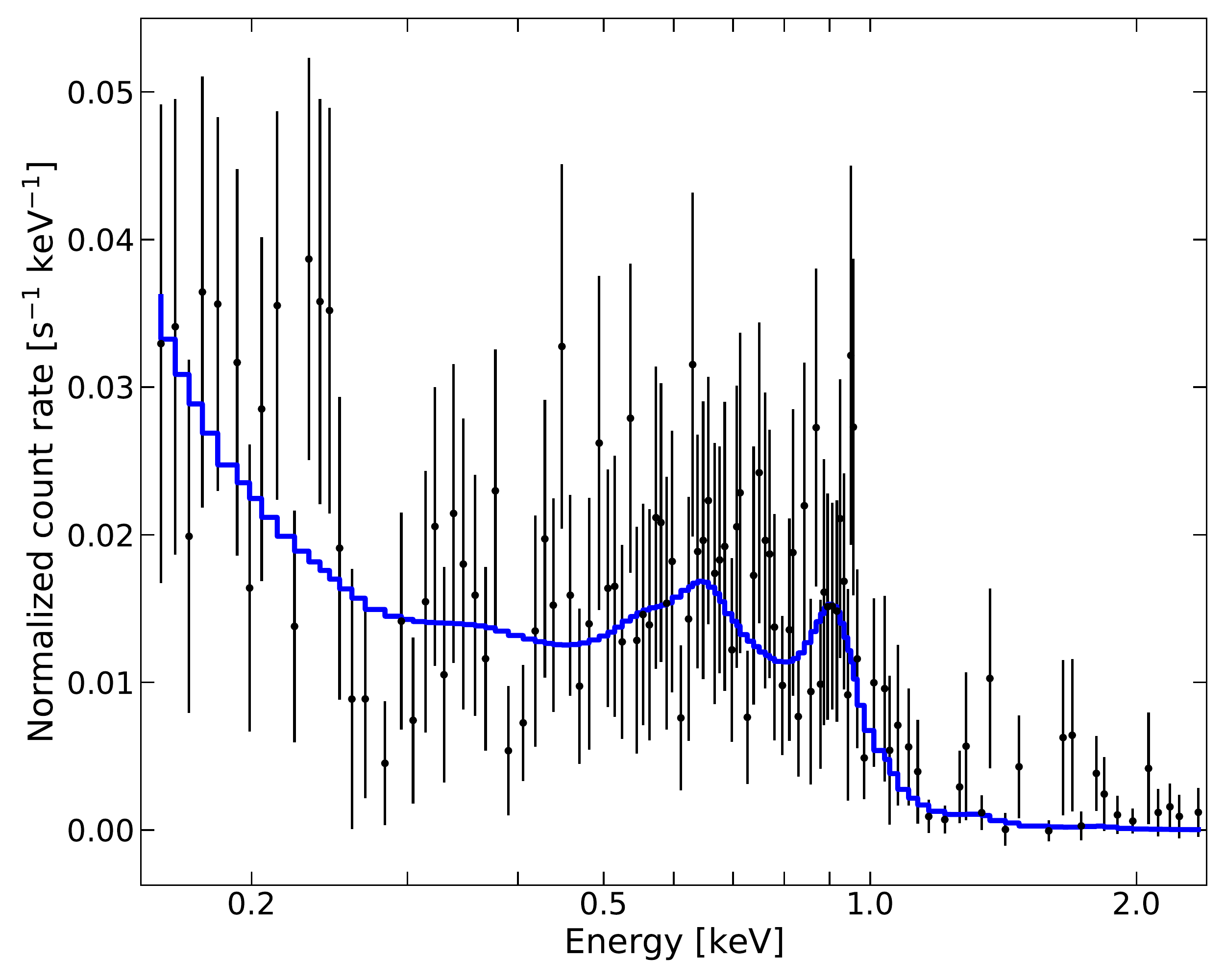}
    \caption{Plot of the best fit for the X-ray spectrum of K2-136 (blue line) binned to 3 counts per bin alongside our \xmm measurements (black data points).}
    \label{fig:vapec-spectrum}
\end{figure}

The X-ray spectrum of K2-136, plotted in Fig. \ref{fig:vapec-spectrum}, is dominated by soft X-rays ($<1$\,keV).
We fitted the spectrum using XSPEC \citep{xspec}, employing the VAPEC model \citep{apec}, which describes the emission spectrum of collisionally-ionized diffuse gas. We allowed the abundances of common elements with strong lines in our energy range to vary (N, O, Ne, and Fe) while keeping the abundances of other elements fixed at Solar values \citep{abund-aspl}.
These elements allow us to characterize the first ionization potential (FIP) bias on the star \citep[following e.g.][]{brinkman-2001:fip, nordon-2013:fip}.

To account for interstellar absorption we used the TBABS model \citep{tbabs} with a fixed hydrogen column density of $n_{\rm H} \approx 10^{18}\rm\,cm^{-2}$, as estimated for the Hyades by \citet{redfield-linksy-2001}.
Given the relatively low number of counts, we binned the spectrum to a minimum of one count per bin and fitted using the Cash statistic \citep{cash-1979}.
Parameter uncertainties were estimated using a Markov chain Monte Carlo (MCMC) method with 20 walkers, 10$^4$ steps and a burn-in of 2000 steps using the algorithm of \citet{gw-mcmc}.

We found that at least three temperature components were required for the model to account for the main features and shape of the spectrum, particularly the rise in flux at softer energies.
The three temperatures components had best fit energies
$kT_1 = 0.11^{+0.03}_{-0.02}$ keV,
$kT_2 = 0.22^{+0.02}_{-0.06}$ keV, and
$kT_3 = 0.80^{+0.04}_{-0.01}$ keV.
Best fitting abundances were:
$2.50_{-0.16}^{+0.48}$ for N;
$1.01_{-0.32}^{+0.80}$ for O;
$2.28_{-1.32}^{+0.26}$ for Ne; and 
$0.53_{-0.06}^{+0.13}$ for Fe (relative to Solar).
The resulting X-ray flux is $F_\text{X} = 2.56 ^{+0.32}_{-0.21}\,\times10^{-14}$ \xflux in the range 0.15 to 2.4 keV.

The flux at energies down to 0.1 keV was estimated by extrapolating our three-temperature model, resulting in a flux $F_\text{X} = 2.80^{+0.43}_{-0.23} \,\times10^{-14}$ \xflux in the range 0.1--2.4\,keV.

The integrated X-ray fluxes in different energy ranges are presented in Table \ref{tab:fluxes} and the three-temperature model is overlaid on Fig. \ref{fig:vapec-spectrum}. The corresponding X-ray luminosity is $L_\text{X} = 1.16 ^{+0.18} _{-0.10} \times 10^{28}$ erg s$^{-1}$ in the band 0.1 -- 2.4\,keV, and thus the X-ray activity is $L_\text{X} / L_\text{bol} = 1.77 ^{+0.40} _{-0.46} \times 10^{-5}$. Furthermore, using the empirical relations by \citet{king-2018}, we estimated an EUV luminosity $L_\text{EUV} = 2.31 ^{+0.28} _{-0.23} \times 10^{28}$ erg s$^{-1}$ at energies $0.0136$ to $0.1$\,keV.

\subsubsection{FIP bias}\label{sec:fip}

Stellar coronal abundances have been observed to be biased according to first ionization potential \citep[FIP;][]{fip-feldman-1992, fip-laming-2015}.
For solar-type stars, elements with higher FIP (e.g.\ O, Ne) are typically present in the corona with lower abundances than lower FIP elements (e.g.\ Fe) in comparison to photospheric abundances.
This is quantified as $F_\text{bias} = \log_{10}{(\text{X}/\text{Fe})}_\text{cor} - \log_{10}{(\text{X}/\text{Fe})}_\text{phot}$ for an element $X$ in comparison to iron \citep{WoodLinsky10:fip-bias, WoodLaming12:fip-bias}.
We find a coronal Ne/Fe value of $4.30^{+1.16}_{-2.54}$, and hence a positive FIP bias $F_\text{bias} = 0.63^{+0.07}_{-0.27}$ for Ne, which suggests a strong {\em inverse} FIP effect, which is characteristic of young active stars \citep{audard-2003:fip, telleschi-2005:fip} as well as late K-dwarfs and M-type stars \citep{gudel-2007:fip, WoodLaming12:fip-bias, fip-laming-2015}.

\subsection{Stellar companion}

\citet{ciardi-2018} reported the presence of a stellar-mass companion 0.7\,arcsec away from the main star using high contrast imaging at the Keck Observatory. Assuming the same distance, this translates to a projected separation of about 40\,AU. Since the spatial resolution (half energy width) of the \xmm telescope EPIC-pn detector is 15\,arcsec \citep{epic-pn}, the candidate companion is unresolved from the primary star in our observations. \citet{ciardi-2018} provided measured magnitudes $J=14.1\pm0.1$, $H=13.47\pm0.04$, and $K_\text{s}=13.03\pm0.03$, and estimated it to be a M7/8V dwarf. Based on its colour index $\text{J}-\text{H}=0.63\pm0.11$, we interpolated with the stellar parameter sequences by \citet{boltable} and obtained a bolometric luminosity of $L_\text{bol}=(1.77\pm0.70)\times10^{30}$ erg s$^{-1}$.

Assuming that this object is indeed an M-dwarf close to K2-136 and a Hyades member, and considering that very low mass stars can remain X-ray saturated for several Gyr \citep{johnstone-2021}, we estimated its maximum contribution to our measured flux using the saturated X-ray activity of $\log_{10} L_\text{X}/L_\text{bol}=-3.13\pm0.08$ determined by \citet{wright-2011}. Its estimated X-ray luminosity is thus $L_\text{X}=(1.3\pm0.6)\times10^{27}$ erg s$^{-1}$ in the band 0.1 -- 2.4\,keV, with a flux that would contribute roughly 5--15\% of the flux in our measurement.  
Since this estimated flux from the companion is comparable to the $1\sigma$ uncertainty in our measurement, 
we chose not to account for the X-ray flux from the stellar companion in our analysis.


\begin{table}
	\centering
	\caption{X-ray fluxes integrated from our three-temperature VAPEC model binned to 1 count per bin.}
	\begin{threeparttable}
	\begin{tabular}{ccc}
		\hline
		\hline
		Energy range & X-ray flux & Label / telescope \\
		(keV) & ($10^{-14}$ \xflux ) & \\
		\hline
		0.15 - 2.40            & $2.56 ^{+0.32} _{-0.21}$  & X-ray, \xmm  \\
		0.10 - 2.40\tnote{a}   & $2.80 ^{+0.43} _{-0.23}$  & X-ray, \textit{ROSAT} \\
		0.0136 - 0.10\tnote{b} & $5.56 ^{+0.68} _{-0.55}$  & EUV          \\
		\hline
	\end{tabular}
	\begin{tablenotes}
        \item[a] Extrapolated from \xmm range.
        \item[b] Calculated from the scaling relation of \citet{king-2018}.
    \end{tablenotes}
	\label{tab:fluxes}
	\end{threeparttable}
\end{table}

\section{Evolution modelling}\label{sec:sim}

In order to simulate the evaporation histories of the planets, we evolved their current states back and forward in time, for which we introduce the \code{photoevolver} \footnote{The code is available on GitHub at \url{https://github.com/jorgefz/photoevolver}} code. This simulation is built upon three main components:
(1) a description of the stellar XUV emission history, which provides the X-ray luminosity of the star at each point in time (Sect.\,\ref{sec:predicted-xuv});
(2) a formulation for the atmospheric mass loss, which will take the incident XUV flux at the planet and translate it into mass lost (Sect.\,\ref{sec:massloss}); and
(3) an envelope structure formulation, which links the envelope mass to its size and describes how the envelope radius responds to mass loss (Sect.\,\ref{sec:struct}).
For each of these three components we explore a range of published models. We adopt and compare three envelope structure models from \citet{lopez-fortney-2014}, \citet{chen-rogers-2016} and \citet{owen-wu-2017}, as well as four mass loss models from \citet{erkaev-2007}, \citet{salz-2016}, \citet{gupta19}, and \citet{kubyshkina-2018}.
In some cases we take equations from those models, in others we use code provided by the authors, and in one case have translated the code into the C programming language to reduce computation time. 

At each simulation time step, the XUV flux is drawn from stellar tracks (Sect.\,\ref{sec:predicted-xuv}) and used to calculate the mass loss rate (Sect.\,\ref{sec:massloss}) and update the envelope mass fraction. Then, the radius is recalculated using the structure formulation (Sect.\,\ref{sec:struct}).

We first solved for the current internal structure of each planet in the system, calculating what fractions of the planetary mass and radius correspond to the core and the gaseous envelope. We then evolved planets that retain a present-day envelope backwards in time from their current age of 700\,Myr to 10\,Myr with 0.1\,Myr time steps. Ten million years is the age at which the protoplanetary disc is fully dissipated \citep{disk-dispersal-10myr} and any boil-off phase has completed \citep{lammer-2016-boiloff, fossati-2017-boiloff, owen-wu-2016:boil-off}, and we refer to this as the planet's initial state for our simulations. 
We also evolve these planets forward to 5 Gyr with 1 Myr time steps. Finally, in order to ensure these evaporation histories are consistent across all planets in the system, we evolved the planets with no present-day envelope forward in time from 10 Myr using a range of initial envelope mass fractions, ensuring that the envelopes are fully stripped within 700 Myr. 
We considered the envelope lost if the envelope mass fraction $\text{X}_\text{env} = \text{M}_\text{env} / \text{M}_\text{core} $ fell below $0.01$\%.
We choose this limit because, although such a planet might be stable \citep{Misener21:residual-envelopes}, our envelope models are not rated below this limit, and the envelope would contribute only a small fraction of the planet's radius. Furthermore, we found that our evolution code removes such envelopes within a few time steps in any case.

\begin{figure}
    \includegraphics[width=\columnwidth]{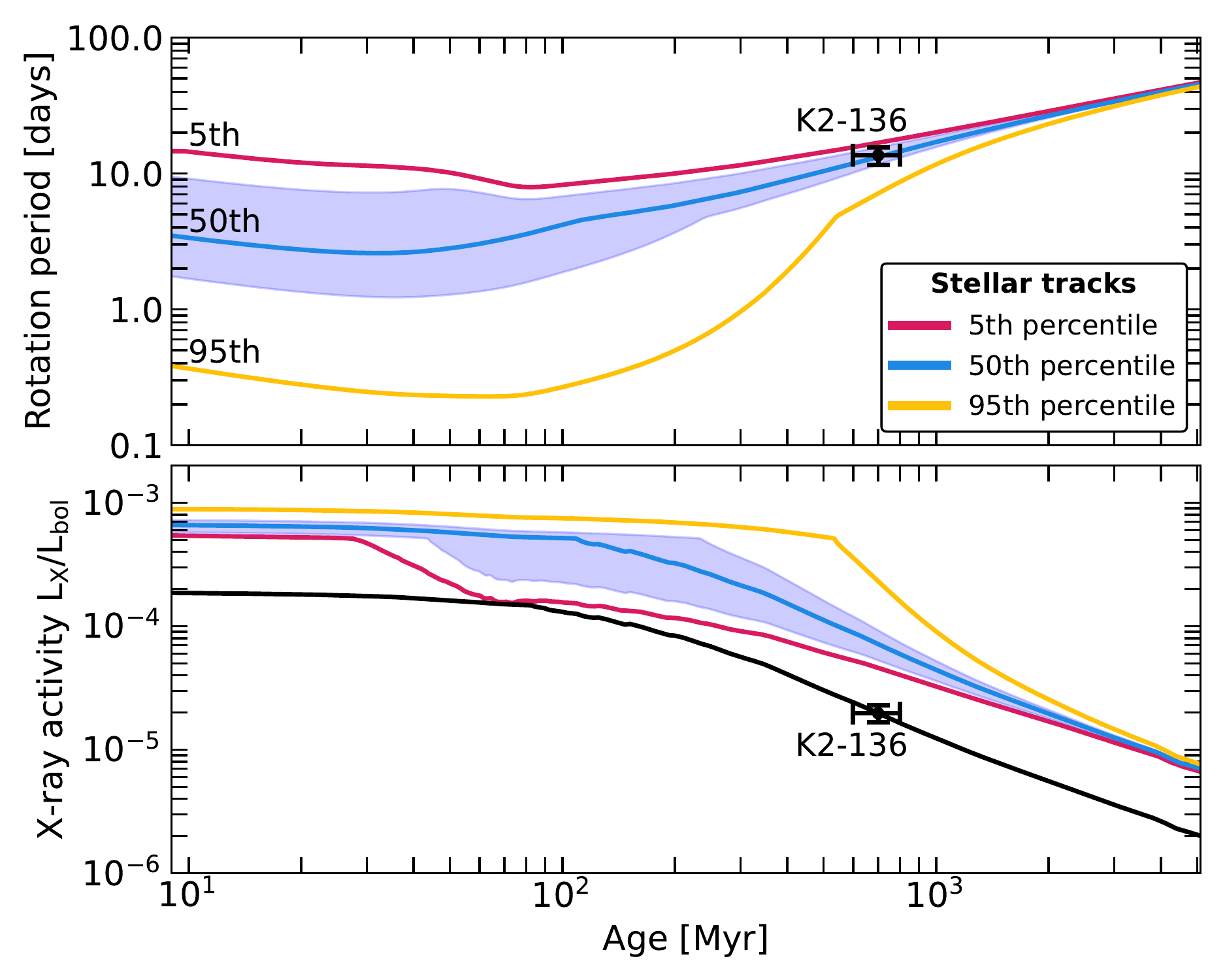}
    \caption{Simulated time evolution of the rotation period (top panel) and the corresponding X-ray activity (bottom panel) of a star of mass $0.69\,M_\odot$ following the models by \citet{johnstone-2021}. The diversity of rotation periods with which a star of this mass starts out is represented with three rotational evolution tracks at the 5th (red), 50th (blue), and 95th (yellow) percentiles. The measured values for the period and X-ray activity of K2-136 are plotted as black points with error bars. The area shaded in blue represents the possible rotational pasts of K2-136 given the uncertainty in its measured period. The black line on the bottom panel corresponds to the model tracks that fit the star based on its mass, period, and age, but scaled to match our X-ray measurement.}
    \label{fig:rotation-activity-evolution}
\end{figure}

\begin{figure}
    \includegraphics[width=\columnwidth]{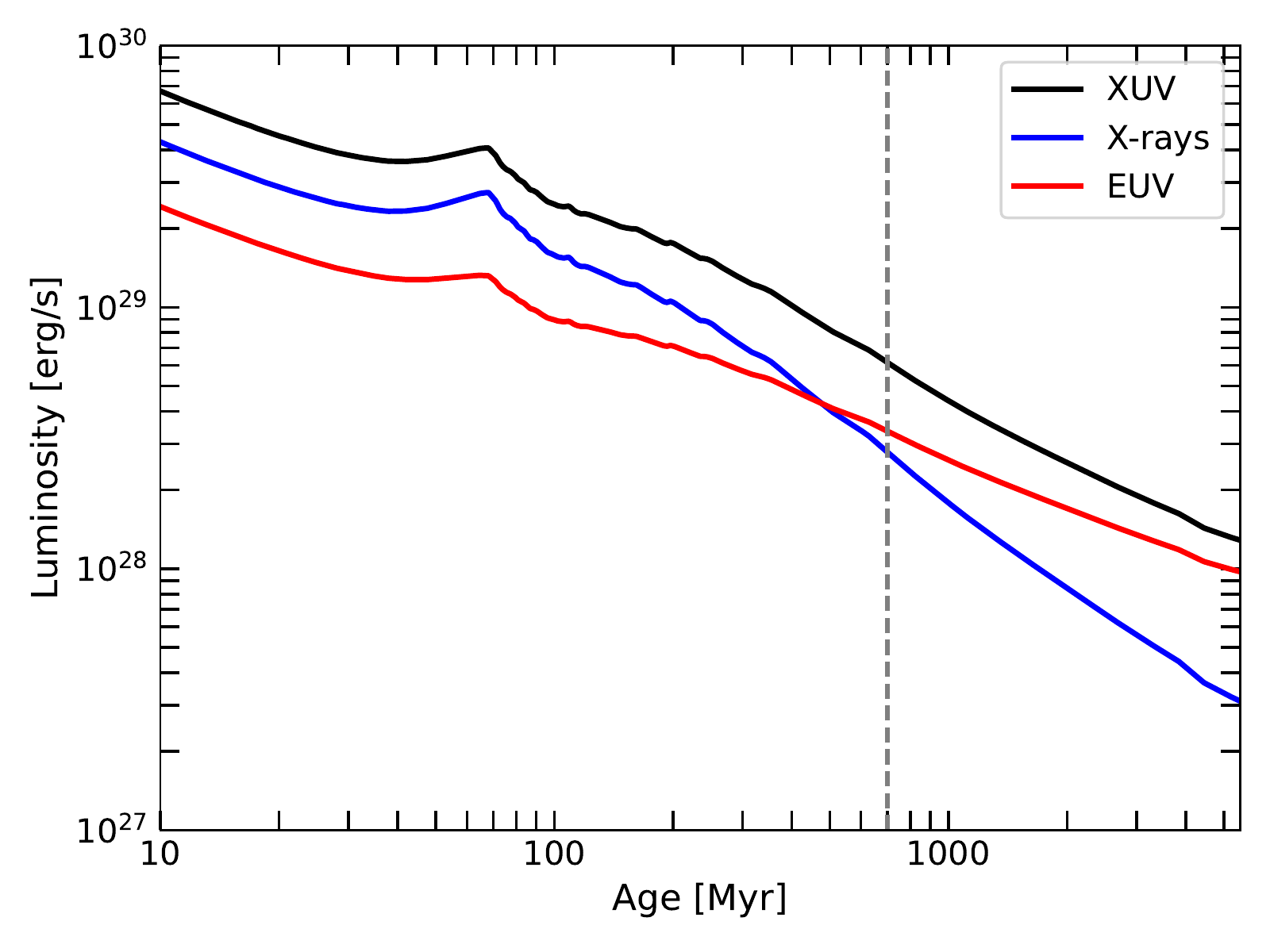}
    \caption{X-ray (blue), EUV (red), and the combined XUV (black) luminosity evolution tracks of K2-136 following the rotational models by \citet{johnstone-2021} for X-rays and the scaling relations by \citet{king-2018} for EUV. The current age of the system, 700 Myr, is plotted as a dashed grey line.}
    \label{fig:xray-vs-euv}
\end{figure}

\subsection{Stellar XUV history}\label{sec:predicted-xuv}

We adopt the stellar rotation description of \citet{johnstone-2021}, who modeled stellar rotational evolution by describing the convective and radiative regions as two solid shells that rotate independently, taking into account different angular momentum transport mechanisms. They combine this description with distributions of rotation periods as a function of stellar mass and age estimated from observations of nearby open clusters. The rotational evolution models for a star of the mass of K2-136 are shown in Fig.\,\ref{fig:rotation-activity-evolution}.

The age of the Hyades cluster (and thus of K2-136) is well constrained (see Table\,\ref{tab:star_params}) and we adopt a value of 700\,Myr.
In addition, \citet{livingstone-2018} measured a rotation period of $13.6^{+2.2}_{-1.5}$ days, in agreement with \citet{mann-2017}, who measured $15.0\pm1.0$\,days, and \citet{ciardi-2018} who identified a spin period of $13.8\pm1.0$\,days based on the periodogram of the K2 lightcurve. This period matches the median (50th percentile) of the model's distribution (Fig.\,\ref{fig:rotation-activity-evolution}), although a range of rotational pasts are possible based on the error on the period, with early-age periods at 10 Myr spanning $1.7$ to $9.2$ days. Nevertheless, we adopt the stellar tracks that fit the star's current mass, age, and rotation period.

The X-ray activity $L_\text{X} / L_\text{bol}$ predicted by this model is about 3.5 times higher than our measured value (Fig.\,\ref{fig:rotation-activity-evolution} and Sect.\,\ref{sec:xray}). This is mainly due to the predicted X-ray luminosity, which is 2.7 times stronger than our \xmm measurement \citep[the predicted bolometric luminosity is only 22\% greater and close to the 20\% uncertainty estimated by][]{livingstone-2018}. This discrepancy is not surprising, given the spread in X-ray emission of about one order of magnitude on the rotation-activity relation \citep[e.g.][]{wright-2011}. The origin of this spread is not clear, although
short timescale
X-ray variability has been proposed as a explanation \citep[e.g.][]{johnstone-2021}. If this were the case, we would expect these variations to average out over time and thus the current X-ray luminosity of K2-136 might be explained by a temporary low-activity season. However, if stars follow intrinsic X-ray evolution tracks and hence the X-ray activity of K2-136 is consistently lower than the model, this would result in lower mass loss rates on its planets. We thus implement two parallel analyses: one with the predicted X-ray luminosity track for K2-136, and another with a low-luminosity track 2.7 times fainter that fits our measurements of K2-136 at its current age of 700 Myr. In Sect.\,\ref{sec:hyades} we compare the X-ray emission of K2-136 with similar Hyades members in order to assess whether the measured or modelled XUV fluxes are the most typical. 

We estimate the corresponding EUV emission for the two X-ray histories using the scaling relations by \citet{king-2018}, who derive empirical relations between X-ray (0.1--2.4\,keV) and EUV (0.036--0.1\,keV) fluxes using Solar \textit{TIMED/SEE} data.
Furthermore, \citet{king-2021} find that the EUV luminosity of a star declines more slowly than X-rays, with EUV irradiation dominating in later ages, which may affect the evolution of planets in Gyr timescales. On Fig. \ref{fig:xray-vs-euv}, we plot both the X-ray and EUV luminosity evolution of K2-136 following this prescription, together with the combined XUV track. Indeed, the X-rays start out twice as strong as EUV at 10 Myr, then become equal at around 480 Myr, and by the age of 5 Gyr the EUV emission ends up three times stronger than X-rays (Fig.\,\ref{fig:xray-vs-euv}).

\begin{figure}
    \includegraphics[width=\columnwidth]{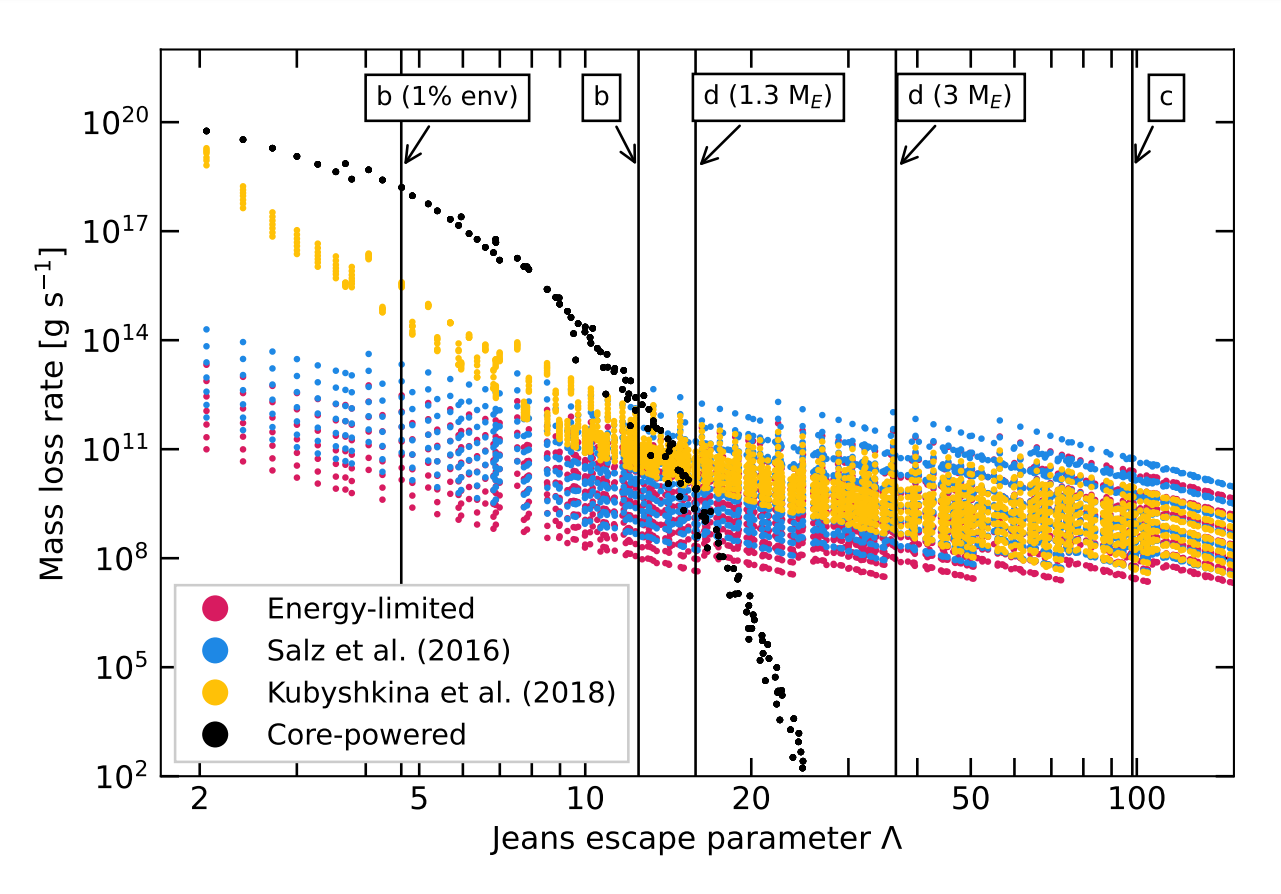}
    \caption{Mass loss rates from the formulations described in Sect. \ref{sec:massloss} as a function of the Jeans escape parameter, following \citet{kubyshkina-fossati-2021}. The mass loss rates are calculated from a grid of parameters that cover the expected incident XUV fluxes as well as masses and radii of the three planets across their evaporation history. The Jeans parameters of the K2-136 planets are shown as vertical lines. For planet b, we show it at its current position, and with a 1\% envelope added (representative of its possible past). For planet d, we show its position for the two possible masses we consider (at 1.3\,M$_\text{E}$ with envelope mass fraction of 0.2\%, and at 3\,M$_\text{E}$ with 0.05\%).
    }
    \label{fig:massloss-jeans}
\end{figure}

\begin{figure*} 
    \includegraphics[width=\textwidth]{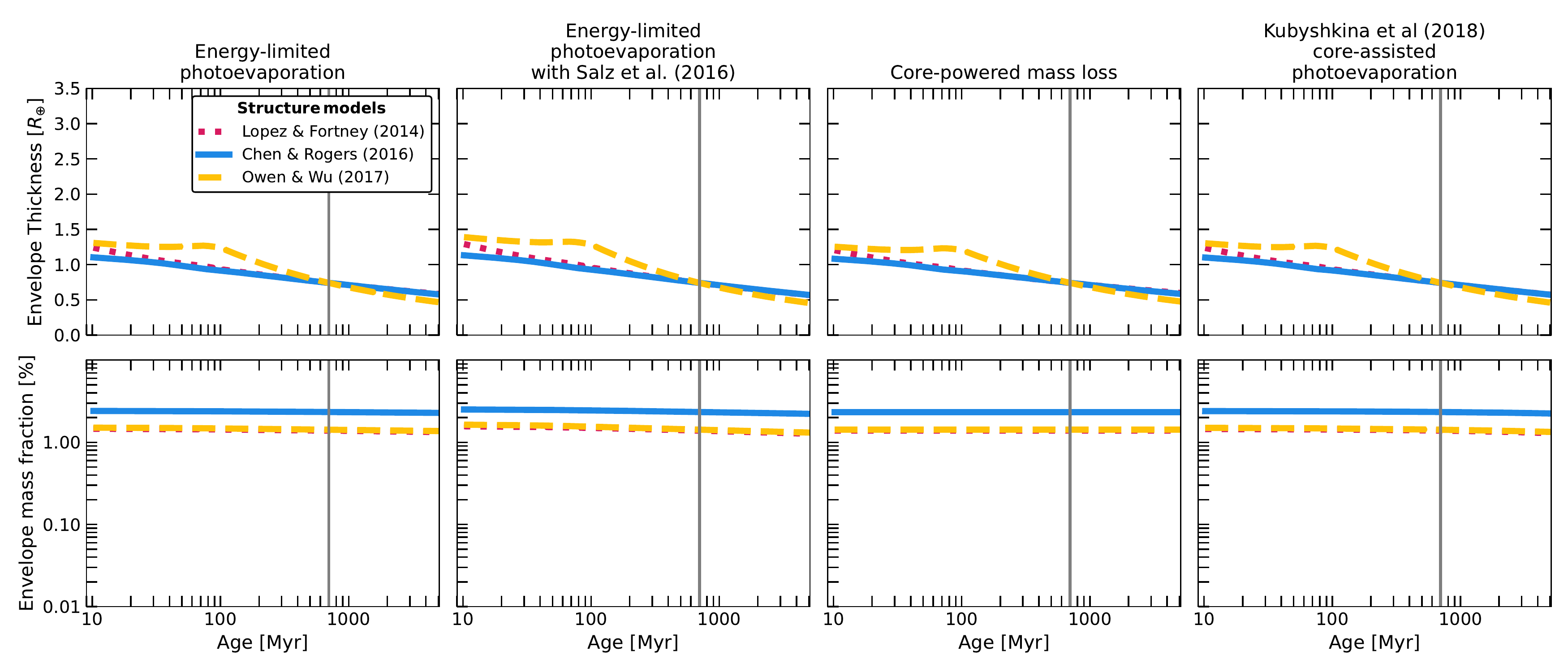}
    \caption{Past and future structure evolution of the planet K2-136\,c, for both its thickness $\text{R}_\text{env}$ (top panels) and its envelope mass fraction $M_{\text{env}}/M_{\text{core}}$ (bottom panels). Each column uses a different mass loss formulation which are explained in detail in Sect.\,\ref{sec:massloss}; from left to right: (1) standard energy-limited photoevaporation, (2) energy-limited photoevaporation enhanced by \citet{salz-2016}, (3) the core-powered mass loss model by \citet{core-powered-mloss}, and (4) the hydrodynamic core-assisted photoevaporation model by \citet{kubyshkina-2018}. The different coloured lines indicate each of the structure models used in the simulation: \citet{lopez-fortney-2014} as a red dotted line, \citet{chen-rogers-2016} as a blue solid line, and \citet{owen-wu-2017} as a yellow dashed line. The grey vertical line indicates the planet's current age. The axes are the same as for K2-136\,d in Fig.\,\ref{fig:evo-d-light}.}
    \label{fig:evo-c}
\end{figure*}

\subsection{Atmospheric mass loss}\label{sec:massloss}

Atmospheric loss models estimate the mass loss rate from a planet's atmosphere. A commonly used X-ray based approximation is the energy-limited approach, in which high-energy radiation incident on the planet is converted to the work needed to bring an atom from the upper atmosphere to the Roche lobe, where it escapes the planet's gravity \citep{watson-1981, lecavelier-des-etangs-2007, erkaev-2007}. The energy-limited mass loss rate is given by 
\begin{equation}\label{eqn:energy-limited}
    \dot{M} = \frac{\pi \eta \beta^2 F_\text{XUV} R^{3}_\text{p}}{G K M_\text{p}},
\end{equation}
where $M_\text{p}$ and $R_\text{p}$ are the mass and radius of the planet, respectively, $F_\text{XUV}$ is the incident X-ray and EUV flux onto the planet, $\eta$ is the energy efficiency of the process, $\beta = R_\text{XUV} / R_\text{p}$ where $R_\text{XUV}$ is the radius at which the atmosphere becomes optically thick for XUV wavelengths; and $K$ is the factor that accounts for the potential difference between the surface for the planet and the Roche lobe, which is given by 
\begin{equation}\label{eqn:roche-lobe-factor}
    K = 1 - \frac{3}{2\xi} + \frac{1}{2\xi^3},
\end{equation}
where $\xi$ is given by
\begin{equation}
    \xi = \frac{R_\text{Roche}}{R_\text{p}} \approx \frac{a}{R_\text{p}} \left( \frac{M_\text{p}}{3M_*} \right)^{1/3},
\end{equation}
and $R_\text{Roche}$ is the Roche radius, $a$ is the semi-major axis of the planet's orbit, and $M_*$ is the mass of the host star \citep{erkaev-2007}.

This approach is attractively simple, although it does mask more complex physics behind the efficiency ($\eta$) and XUV radius ($\beta$) parameters, which are difficult to determine. In the first instance, we have assumed an efficiency of 15\% \citep{lammer-2009, jackson-2012, shematovich-2014, king-2019} as well as an XUV radius $R_\text{XUV} = R_\text{p}$ ($\beta$ = 1), which is a physical lower limit. However, observations of high-energy transits have found XUV radii to be larger than optical radii \citep{linsky-2010, poppenhaeger-2013-xray-transit, bourrier-2016}, and hence we also adopt the formulation by \citet{salz-2016}, who provide estimates for XUV radii using a grid of planetary gravitational potentials and incident XUV fluxes generated using hydrodynamic simulations.

\citet{kubyshkina-2018} introduced an upper atmosphere hydrodynamic model based on the work by \citet{johnstone-2015-massloss-model}. They compute radial profiles for atmospheric velocity, temperature, and density, and hence estimate the atmospheric escape rate at the Roche lobe by taking into account high energy radiation at two wavelengths (EUV at 60\,nm and X-rays at 5\,nm) as well as internal core heating. The model deviates from the energy-limited approach for highly-irradiated low-gravity planets, where it predicts greater mass loss rates. We adopt the updated model grid and interpolation routine provided by \citet{kubyshkina-fossati-2021}.

Alternative formulations that do not rely on XUV irradiation have also been suggested. One such example is core-powered mass loss \citep{core-powered-mloss}, which draws the energy to evaporate envelopes from the core’s internal thermal luminosity. This formulation has also been shown to replicate the observed exoplanet radius-period distribution \citep{ginzburg18,gupta19}, and operates on timescales of Gyr. \citet{king-2021}, however, have shown that significant EUV irradiation  also continues on Gyr timescales, and thus the timescale of mass loss alone cannot easily be used to distinguish between the photoevaporative and core-powered mechanisms. 

We run a set of simulations that combine each structure formulation described in Sect. \ref{sec:struct} with the following mass loss descriptions:
(1) energy-limited with 15\% efficiency and $R_\text{XUV} = R_\text{p}$, which we refer as `standard` hereafter, 
(2) energy-limited with 15\% efficiency and $R_\text{XUV}$ described by \citet{salz-2016}, 
(3) core-powered mass loss \citep{core-powered-mloss}, and 
(4) hydrodynamic simulations of \citet{kubyshkina-2018}, which reproduce the core-powered and photoevaporative mass loss regimes.

In Fig. \ref{fig:massloss-jeans} we compare these mass loss formulations as a function of the Jeans escape parameter, which quantifies how vulnerable a gaseous atmosphere is to escape due to thermal escape mechanisms. This parameter is defined as
\begin{equation}\label{eqn:jeans}
    \Lambda = \frac{G M_\text{p} m_\text{H}}{k_\text{B} R_\text{p} T_\text{eq}},
\end{equation}
where $M_\text{p}$ and $R_\text{p}$ are the planet's mass and radius, respectively, $T_\text{eq}$ is the equilibrium temperature, $m_\text{H}$ is the mass of a hydrogen atom, and $k_\text{B}$ is the Boltzmann constant. Following \citet{kubyshkina-fossati-2021}, two mass loss regimes can be identified. Atmospheres in the low-$\Lambda$ regime are loosely bound and are more susceptible to mass loss; this regime contains planets that are highly irradiated and/or low gravity. The high-$\Lambda$ regime, on the other hand, applies to massive planets with tightly bound envelopes.
In practice, individual planets will tend to evolve towards higher $\Lambda$ (right) and lower mass loss rates (down) as envelopes cool and shrink, and as the stellar XUV emission declines. 

As expected, the energy-limited formulation is not a strong function of Jeans parameter (Fig. \ref{fig:massloss-jeans}), with most points lying on the same range of mass loss rates across $\Lambda$, as mass loss scales linearly with flux and inversely with potential. Furthermore, both the \citet{kubyshkina-2018} and the core-powered formulations predict mass loss rates several orders of magnitude greater than energy-limited on planets with low Jeans parameter, but differ the most on the high-$\Lambda$ regime, where \citet{kubyshkina-2018} tends to maintain levels similar to energy-limited in most cases whereas core-powered drops steeply and the escape rate becomes negligible for $\Lambda>20$.

\subsection{Envelope structure models}\label{sec:struct}

Models typically describe the gaseous planetary envelope  as having two layers: an inner adiabatic and convective layer, and an outer radiative layer that is isothermal at the equilibrium temperature. Sources of heat act to inflate the envelope, and heat loss causes it to shrink. In our simulations, we consider three envelope structure models from the studies of \citet{lopez-fortney-2014}, \citet{chen-rogers-2016}, and \citet{owen-wu-2017}. The envelope structure model by \citet{lopez-fortney-2014}
provides the envelope thickness as a function of planet mass, envelope mass fraction, bolometric flux, and age.
It considers the core entropy and radioactive elements within the planet as well as the stellar bolometric flux as sources of heat, and heat loss occurs through wavelength-dependent radiative cooling. The optical radius of the planet is defined by a fixed atmospheric pressure. Additionally, they also use hot-start models, which imply a high starting core entropy. This assumption can result in particularly large initial radii, although it also leads to rapid cooling and envelope shrinking at early ages, making the choice of initial entropy unimportant by the age of 100 Myr \citep{lopez-fortney-2012}.
We adopt the polynomial fit they provide in \citet[][equation 4]{lopez-fortney-2014}, which is an empirical characterisation of a precalculated grid of models covering envelope mass fractions between 0.01\% and 20\%.

The envelope model from \citet{chen-rogers-2016} follows a similar description to \citet{lopez-fortney-2014} but they make use of the 1D stellar evolution code MESA (Modules for Experiments in Stellar Astrophysics) adapted to H/He planetary envelopes. This model also defines the planet radius as the height at a fixed optical depth, in order to avoid issues that arise with small planets with puffy initial states contracting rapidly. In our simulations, we adopt the quadratic fit to a precalculated model grid provided by \citet[][equation 5]{chen-rogers-2016}, which is also valid for envelope mass fractions between 0.01\% and 20\%.

Finally, \citet{owen-wu-2017} introduced a fully analytical approach for calculating the envelope thickness considers the heat transport through the convective-radiative boundary in the atmosphere as well and use it to recreate the radius valley adopting envelope mass fractions from 0.01\% to 60\% and ages from 1 Myr to 3 Gyr. They also provide the code \texttt{EvapMass} which implements their structure model \citep{Owen20:EvapMass}. In this case, we have translated their Python code into the C programming language to shorten the computation time.
This code is included in \texttt{photoevolver}, and
we have ensured that it produces identical results to the original Python code.

\begin{figure*}
    \includegraphics[width=\textwidth]{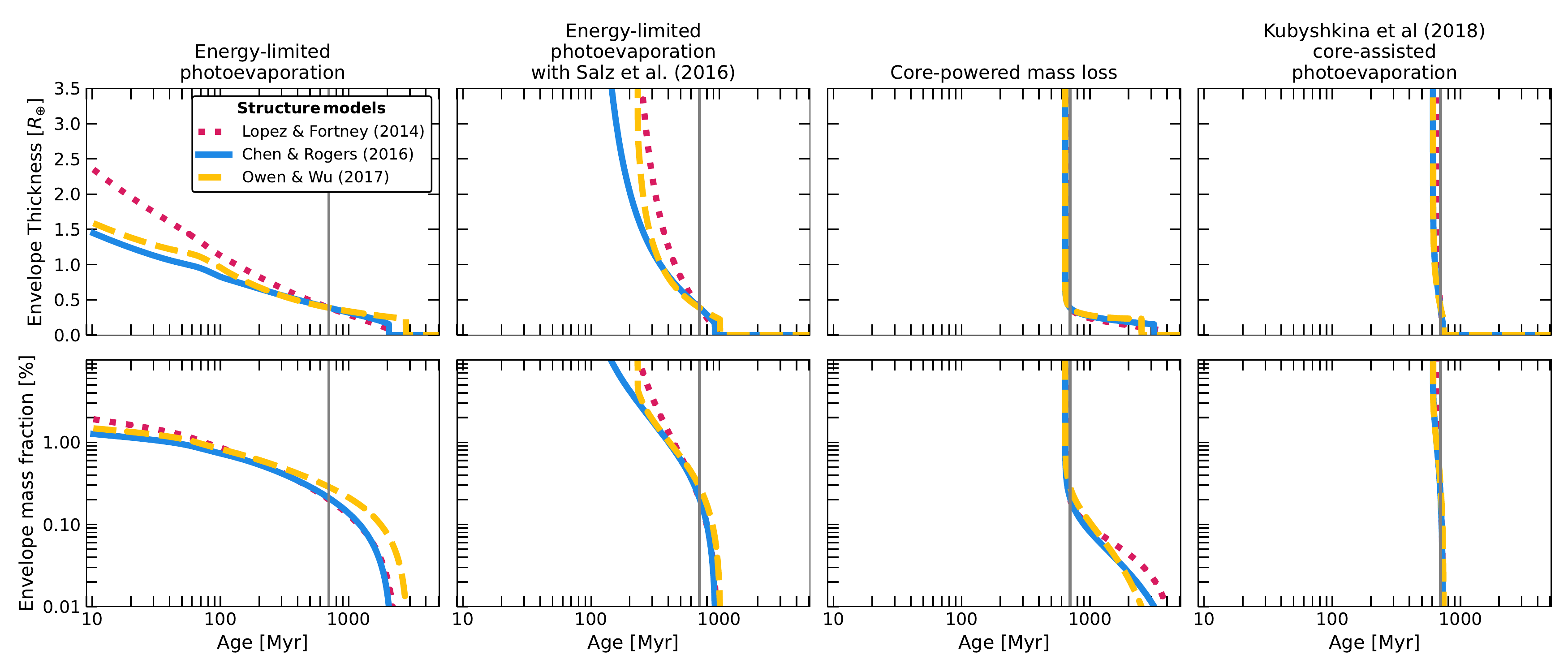}
    \caption{Past and future structure evolution of the planet K2-136\,d with a core mass of 1.3 $\Mearth$, akin to Fig. \ref{fig:evo-c}.}
    \label{fig:evo-d-light}
\end{figure*}

\section{Simulation results}\label{sec:results}

\subsection{Current planet structures and mass loss rates}
\label{sec:current}

For K2-136\,c, we estimate the current envelope mass fraction $M_\text{env}/M_\text{core}$ to be $1.1\%$ with the models of \citet{lopez-fortney-2014} and \citet{owen-wu-2017}, and $1.7\%$ with \citet{chen-rogers-2016}, where $M_\text{env}$ is the mass of the gaseous envelope and  $M_\text{core}$ is the mass of the rocky core. 
For K2-136\,d, using the 1$\sigma$ upper limit on the mass ($<1.3\,\Mearth$; Table\,\ref{tab:planets}), we estimate  an envelope mass fraction of 0.2\% for all three models. As noted in Sect.\,\ref{sec:planets}, the 2$\sigma$ upper limit on the mass of K2-136\,d is consistent with a bare rocky core, with no significant gaseous envelope surviving to the present day.  
The mass constraints for K2-136\,b are less precise and allow a gaseous envelope, but due to its small radius and proximity to its host star we assume it is a rocky core at the present day. 

The current mass loss rates for planets c and d are shown in Table\,\ref{tab:mloss-rates} for each of the mass-loss prescriptions described in Sect.\,\ref{sec:massloss}. 
Due to its lower gravitational potential, K2-136\,d has higher predicted mass loss rates overall than K2-136\,c, despite being more distant from its host star. 
Furthermore, in the case of K2-136\,d, standard energy-limited predicts lower escape rates in comparison to other formulations, whereas for K2-136\,c the energy-limited formulations predict similar escape rates to \citet{kubyshkina-2018}. Those authors argue that the energy-limited formulation underestimates mass loss on very low gravity planets, as it does not account for the contribution of the planet's internal thermal energy in assisting atmospheric escape.
Finally, in the case of planet c, core-powered mass loss predicts a negligible escape rate due to the deeper gravitational potential.

\begin{table}
	\centering
	\caption{Current mass loss rates for planets c and d using the mass loss prescriptions described in Sect.\,\ref{sec:massloss}. From left to right: standard energy-limited (EL), enhanced energy-limited according to \citet[S16]{salz-2016}, core-powered mass loss by \citet[GS20]{gupta19}, and the hydrodynamic simulations of \citet[K18]{kubyshkina-2018}.}
	\begin{threeparttable}
	\begin{tabularx}{0.9\columnwidth}{l|XXXX}
		\hline
		\hline
		Planet  & \multicolumn{4}{c}{Mass loss rate ($10^{8}$ g s$^{-1}$)} \\
		        &  EL & S16 & GS20 & K18 \\
		\hline
		K2-136\,c          & 8.7 & 18.3 & 0.0\tnote{b}  & 10.4  \\
		K2-136\,d\tnote{a} & 7.7 & 32.2 & 36.5 & 237 \\
		\hline
	\end{tabularx}
	\begin{tablenotes}
        \item[a] Using core mass $M_\text{core} = 1.3\,\Mearth$.
        \item[b] Mass loss rate below $10^3$ g s $^{-1}$ (negligible).
    \end{tablenotes}
	\label{tab:mloss-rates}
	\end{threeparttable}
\end{table}

\subsection{Evaporation histories}
\label{sec:histories}

\subsubsection{K2-136\,c}

Firstly, we find that K2-136\,c has remained relatively unchanged during its lifetime, as shown in Fig. \ref{fig:evo-c}, regardless of the choice of envelope and mass loss prescriptions. This is  due to its high mass of $18.1\pm1.9\,\Mearth$ and thus deep gravitational potential which inhibits atmospheric loss. 
All scenarios predict an initial planet size at 10\,Myr of $R_\text{p} = 3.5\,\Rearth$, following any boil-off phase \citep[e.g.][]{owen-wu-2016:boil-off}. Subsequently, the planet shrinks primarily due to the thermal cooling and contraction of its envelope. Mass loss is minimal, with an almost constant envelope mass fraction throughout its lifetime. We also find that this planet will not lose its envelope in the next several Gyr.
The bump apparent in the envelope model by \citet{owen-wu-2017} model in Fig.\,\ref{fig:evo-c} is a feature of their formulation. They argue that boil-off early in the planet's evolution leads to a rapid cooling and loss of the primordial gaseous envelope, affecting the thermal evolution of the envelope for the first 100 Myr. They address this phenomenon by halting age-dependent thermal evolution for this period of time. 

\begin{figure*}
    \includegraphics[width=\textwidth]{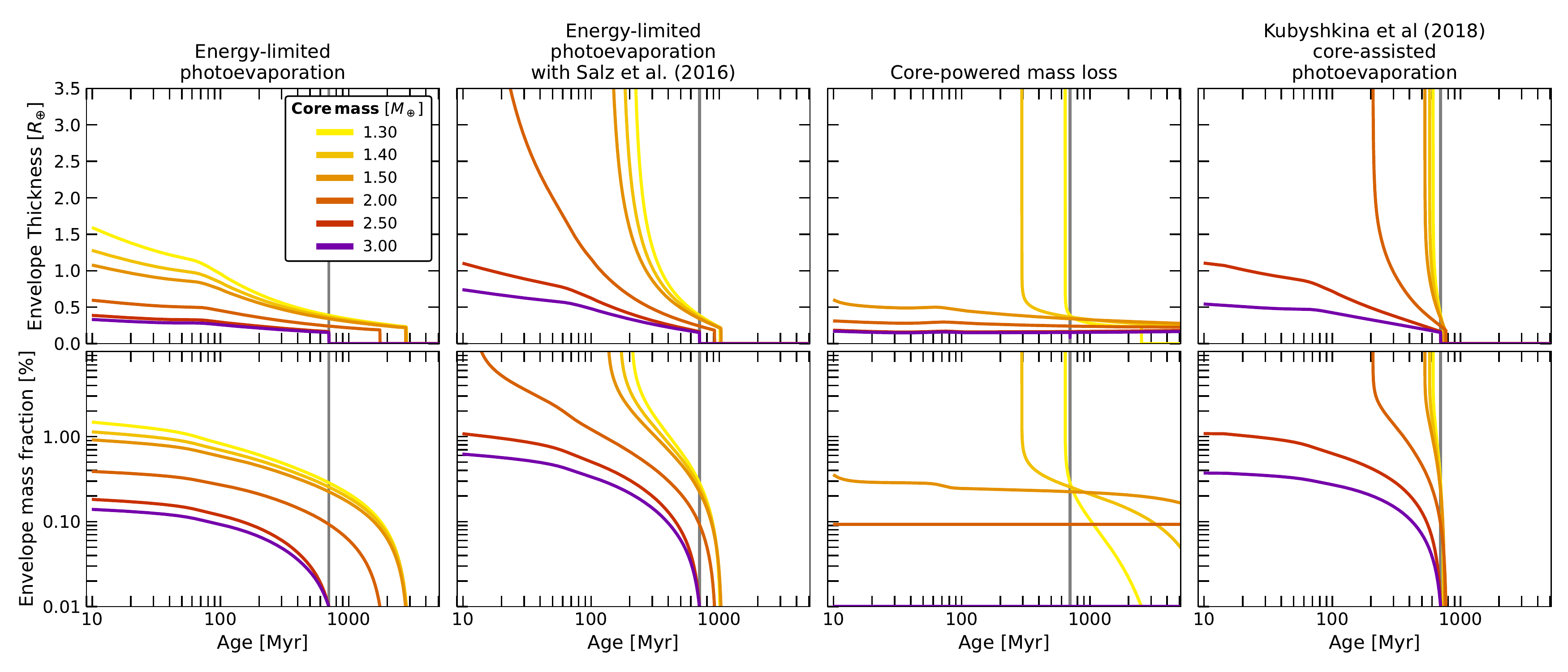}
    \caption{Past and future evolution of the planet K2-136\,d with a range core masses (and radii) from lightest (yellow) to heaviest (purple): 1.3 $\Mearth$ (1.11 $\Rearth$), 1.4 $\Mearth$ (1.14 $\Rearth$), 1.5 $\Mearth$ (1.16 $\Rearth$), 2.0 $\Mearth$ (1.26 $\Rearth$), 2.5 $\Mearth$ (1.34 $\Rearth$), and 3.0 $\Mearth$ (1.42 $\Rearth$). The corresponding solid core radii are calculated using the mass radius relations by \citet{otegi-2020}. The evolution of the envelope thickness $\text{R}_\text{env}$ is shown on the top panels and the corresponding mass fraction on the bottom panels. All tracks make use of the structure formulation by \citet{owen-wu-2017}. A different mass loss model is used for each column, which are described on Setc.\,\ref{sec:massloss}; from left to right: (1) standard energy-limited photoevaporation, (2) energy-limited photoevaporation enhanced by \citet{salz-2016}, (3) the core-powered mass loss model by \citet{core-powered-mloss}, and (4) the hydrodynamic photoevaporation model by \citet{kubyshkina-2018}.}
    \label{fig:evo-d-mcores}
\end{figure*}

\subsubsection{K2-136\,d}
\label{sec:histories-d}

The evolution of K2-136\,d is shown in Fig. \ref{fig:evo-d-light}, in which a core mass of $M_\text{core} = 1.3\,\Mearth$ was used (see Sects.\,\ref{sec:planets}\,\&\,\ref{sec:current}). The planet's envelope experiences significant evolution in both its past and its future.
The standard energy-limited model predicts an initial planet radius of 2.6 to 3.3 $\Rearth$, with initial envelope mass fractions of 1 to 2\%, and complete envelope loss within the next 1 to 2 Gyr.
All other mass loss formulations, however, predict an apparent runaway enlargement during backwards evolution. 
This difference is consistent with the argument by \citet{kubyshkina-2018} that energy-limited formulations underestimate mass loss for low-gravity planets.

Apart from standard energy-limited escape, the models in Fig.\,\ref{fig:evo-d-light} indicate that the planet, with a core mass of $1.3\,\Mearth$, requires an very large initial envelope fraction (more akin to a gas giant with $M_\text{env} \gg M_\text{core}$) in order to still retain part of its envelope by the age of 700 Myr.
However, the steepness of the evolution around the current time shows that models with very large initial envelope fractions would require very fine tuning to reproduce the state of K2-136\,d at its current age.
Instead, we conclude that K2-136\,d is more likely to have a mass significantly in excess of $1.3\,\Mearth$ (which is the $1\,\sigma$ upper limit from HARPS-N, see Sect.\,\ref{sec:planets}). 

We now proceed to find a lower limit to the mass of K2-136\,d, such that the presence of a small envelope (to explain its size and density) is consistent with our simulated evaporation histories. The upper limit of the mass, on the other hand, comes from assuming that K2-136\,d is entirely rocky with a mass of $3.6\,\Mearth$.

In order to constrain the minimum viable mass of K2-136\,d, we study its evaporation history as a function of mass. We apply the same analysis with a range of core masses between $M_\text{p}=1.3\,\Mearth$ (corresponding to $R_\text{core}=1.1\,\Rearth$) to $M_\text{p}=3.0\,\Mearth$ (with $R_\text{core}=1.4\,\Rearth$). We adopt a single structure model for our analysis, \citet{owen-wu-2017}, which predicts intermediate planet sizes in comparison to \citet{lopez-fortney-2014} and \citet{chen-rogers-2016}, and is applicable to ages younger than 100 Myr.
Our results in Fig. \ref{fig:evo-d-mcores} show that the energy-limited model is consistent with the full range of masses and predicts initial planet radii ranging from 1.8 to 2.7 $\Rearth$, with envelope mass fractions of 0.15 to 1.5\%, for the heaviest and lightest cases, respectively.
For the other mass-loss formulations, we define a threshold core mass above which the planet does not grow exponentially in size to earlier times. 

We thus estimate minimum core masses of 1.5 $\Mearth$ with core-powered mass loss \citep{gupta19}, 2.0 $\Mearth$ with enhanced energy-limited escape \citep{salz-2016} and 2.5 $\Mearth$ for the model of \cite{kubyshkina-2018}. 

\citet{salz-2016} and \citet{kubyshkina-2018} generally agree in their results for the 3.0 and 2.5 $\Mearth$ cores, predicting initial radii of 2.0 and 2.4 $\Rearth$ and corresponding envelope mass fractions of 0.5\% and 1.0\% for the two core masses, respectively (Fig.\,\ref{fig:evo-d-mcores}). These results would classify K2-136\,d as a sub-Neptune for at least 100 Myr after formation. These two formulations, however, diverge on their assessment of the 2.0 $\Mearth$ case.
Whilst \citet{kubyshkina-2018} predicts runaway enlargement during backward evolution, \citet{salz-2016} predicts an inflated Saturn-sized planet with radius 9.9 $\Rearth$ and mass fraction 11\% as the initial state.

In contrast, core-powered mass-loss \citep{gupta19} predicts much lower mass loss rates for the higher mass cores, and thus only discards the 1.3 and 1.4 $\Mearth$ core mass scenarios. For its minimum viable mass case of 1.5 $\Mearth$, it predicts a smaller initial planet size of 1.7 $\Rearth$ (mass fraction 0.3\%), which would place the planet at the centre of the radius valley after formation.

Looking at the future evolution of planet d in Fig.\,\ref{fig:evo-d-mcores}, complete envelope loss occurs fairly rapidly with the \citet{salz-2016} and \citet{kubyshkina-2018} models, with both mass-loss prescriptions stripping the envelopes within tens to 200 Myr into the future. Standard energy-limited escape is slower, but still results in complete loss of the envelope within 2\,Gyr. 
Core-powered mass loss predicts complete envelope loss only for the lightest core of 1.3 $\Mearth$ within 2 Gyr.

Finally, we explore the case where K2-136\,d is entirely rocky, with a mass of $3.6\,\Mearth$.
For this planet, we can find an upper limit on the mass of its primordial atmosphere, which should be completely removed by its current age in order to be consistent with photoevaporation.
To do so, we first add an envelope consisting of 0.01\% of its mass (which would be stripped within one simulation time step), and evolve the planet backwards in time using the \citet{kubyshkina-2018} model. This leads to a relatively low upper limit on the initial envelope mass fraction of such a planet of just 0.2\%. This scenario is also in tension with constraints on the total mass of the planet (Sect.\,\ref{sec:planets}).

\subsubsection{K2-136\,b}

For K2-136\,b, the smallest and most close-in of the three planets, we ran our evolution code forward from 10\,Myr using initial envelope mass fractions ranging from 0.1\% to 5\% and assuming a core mass of $1\,\Mearth$. We find all of these initial envelopes are lost within 5 to 30\,Myr even with the standard energy-limited model. All the other mass loss prescriptions are even faster, with envelope loss timescales of\,1-2 Myr. 
These results were expected, given the planet's small size and close proximity to the star, and confirm that any primordial envelope will have been lost early in its evolution.

The initial radius of K2-136\,b can be estimated by mirroring the initial states of the other two planets. An initial envelope mass fraction of about 1.5\%, emulating K2-136\,c, would translate to an initial radius of $2.8\,\Rearth$, using the \citet{owen-wu-2017} structure formulation. Likewise, an initial mass fraction of 0.6\%, following the higher mass scenarios for K2-136\,d, results in an initial radius of $2.1\,\Rearth$. Both cases would qualify the planet as a sub-Neptune at disk dispersal.

\begin{figure*}
    \includegraphics[width=0.75\textwidth]{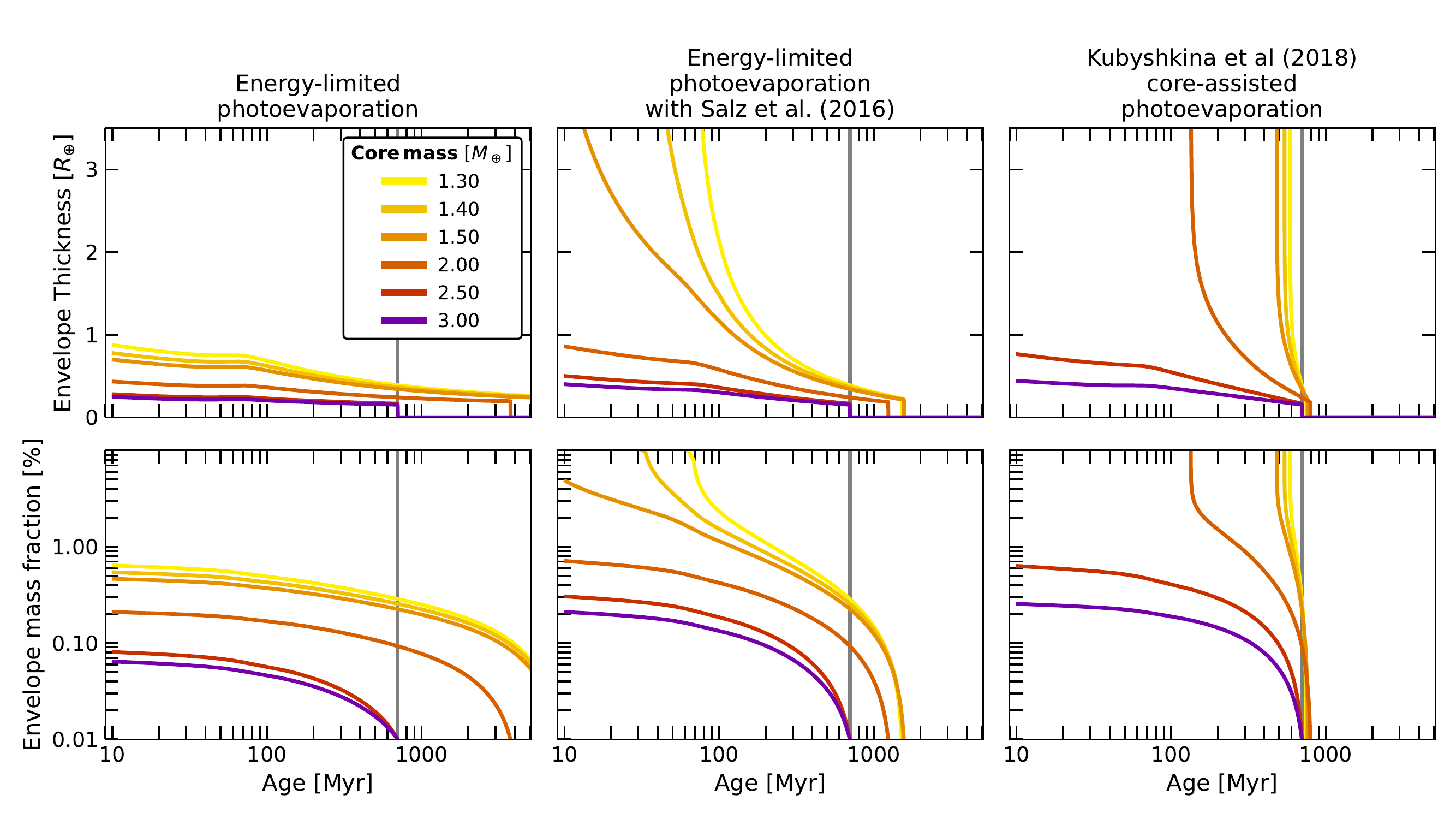}
    \caption{Past and future evolution of the planet K2-136\,d presented in a similar manner to Fig. \ref{fig:evo-d-mcores} but with stellar XUV tracks that fit our measured X-ray luminosity of K2-136. Core-powered mass loss is omitted as it is unaffected by changes in X-ray irradiation.}
    \label{fig:evo-d-mcores-lowxuv}
\end{figure*}

\subsection{Alternative low-level XUV history}
\label{sec:alternative}

We repeated our analysis from Sect.\,\ref{sec:histories} using an alternative low-luminosity XUV track that scales the \citet{johnstone-2021} model to our X-ray flux measured with \xmm\ in Sect.\,\ref{sec:xray-spectrum}. This flux is a factor 2.7 lower than used in Sect.\,\ref{sec:histories}, and we note it is a better match to the X-ray fluxes of similar stars in the Hyades cluster (see Sect.\,\ref{sec:hyades}).

As expected, our results for K2-136\,c are essentially identical, with negligible mass loss and the evolution progressing as a gradual thermal contraction of its envelope. Results for K2-136\,b are also essentially unchanged, with complete loss of any primordial envelope occurring early in the lifetime of the system. 

Our results for K2-136\,d are plotted in Fig. \ref{fig:evo-d-mcores-lowxuv}, excluding core-powered mass loss, which does not depend on high energy irradiation (and is hence unchanged from Fig.\,\ref{fig:evo-d-mcores}). As expected, the lower XUV flux results in reduced mass loss for all of the other three models. 
For standard energy-limited mass loss, the predicted initial radii are roughly 50\% smaller.
Similarly, the \citet{salz-2016} and \citet{kubyshkina-2018} models predict lower initial radii for tracks that do not lead to very rapid evaporation. 
For the \citet{kubyshkina-2018} model, only the 2.5 and 3.0\,$\Rearth$ tracks are consistent with formation as a sub-Neptune, as before, but for the \citet{salz-2016} enhanced energy-limited model, the 1.5\,$\Mearth$ core mass scenario becomes viable, bringing the model closer to consistency with the 1$\sigma$ upper limit on the measured mass (Table\,\ref{tab:planets}). This 1.5\,$\Mearth$ track has an initial envelope mass fraction of 5\% and planet radius of 5.6 $\Rearth$.
A core mass between 1.5 and 2.0\,$\Mearth$ would now mirror the initial envelope mass fraction of K2-136\,c of approximately 1.5\%.

\section{Discussion}\label{sec:discussion}

Based purely on the radii and orbital periods of the planets orbiting K2-136, the system presents a challenge to photoevaporation as an explanation of the radius-period valley. The innermost planet, K2-136\,b, lies below the radius valley, and can easily be explained as the stripped core of a sub-Neptune. The next planet, K2-136\,c, is above the valley and can be understood as having survived the strongest XUV irradiation from the young star. However, the widest-separation planet, K2-136\,d, lies below the radius valley, suggesting it has lost any primordial envelope. The simultaneous presence of planets c and d in the same system, having experienced the same XUV flux history, presents useful constraints on the process of photoevaporative mass loss on multi-planet systems. 

The two key factors controlling photoevaporative mass loss from exoplanets are the XUV radiation environment (and its history) and the response of the planetary envelope to irradiation as a function of core mass. We discuss the XUV radiation environment of the K2-136 planets in Sect.\,\ref{sec:hyades}, in context with its membership of the Hyades cluster. We discuss the response of the K2-136 planetary envelopes to that irradiation in Sect.\,\ref{sec:evaporation}.

\begin{figure*}
    \includegraphics[width=\textwidth]{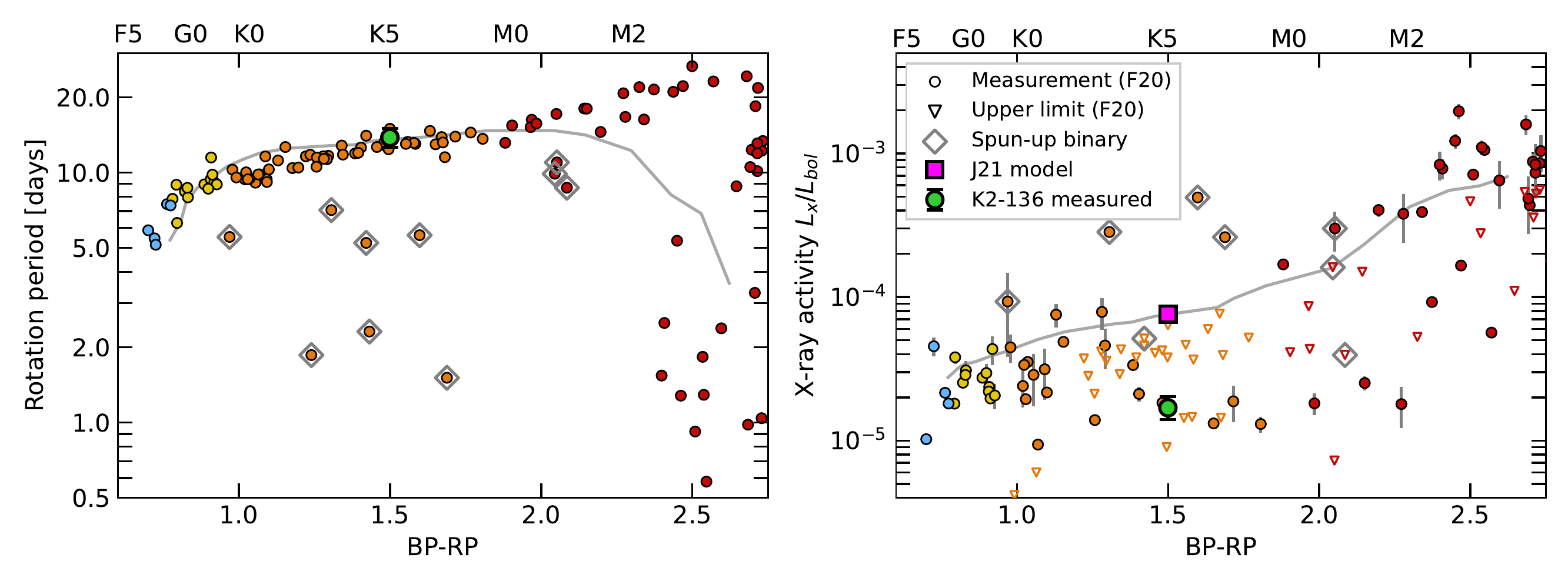}
    \caption{
    \textbf{Left panel:} Hyades stars from \citet[][F20]{freund-2020} plotted on a Gaia BP-RP colour index against rotation period plot. The stars are also colour-coded to spectral type (F stars light blue, G stars yellow, K dwarfs orange, and M dwarfs red). K2-136 is shown as a green circle, and the outliers we identify as spun-up close binaries are marked with diamonds. The grey line represents the mean rotation period as a function of stellar mass for Hyades stars predicted by the model of \citet[][J21]{johnstone-2021}.
    \textbf{Right panel:} Hyades stars from \citet[][F20]{freund-2020} plotted on a Gaia BP-RP colour index against measured X-ray activity $L_\text{X}/L_\text{bol}$ in the energy band $0.1-2.4$\,keV, with markers and symbols similar to the panel on the left.
    X-ray upper limits ($2\sigma$) are shown as triangles. The X-ray activity we determined for K2-136 is shown as a green circle, and the activity predicted by the model of J21 is shown as a magenta square. It can be seen that our measured X-ray flux for K2-136 is a better match to the other Hyades K stars than the flux predicted by J21.}
    \label{fig:hyades}
\end{figure*}

\subsection{K2-136 in context within the Hyades}
\label{sec:hyades}

\citet{freund-2020} compiled a Hyades membership list using Gaia DR2 \citep{gaia-dr2} and collected X-ray flux measurements from 281 Hyades members, 103 of which also have measured rotation periods, and estimated X-ray upper limits for the undetected ones. On Fig.\,\ref{fig:hyades} (left hand panel), we plot the rotation dependence on Gaia BP-RP colour index (the difference between blue and red Gaia magnitudes) for the Hyades, which is indicative of the stellar spectral class. FGK and early M stars present a tight relation between temperature and period that has been observed in other open clusters \citep[e.g.][]{hartman-2009:M37, agueros-2011:praesepe, gillen-2020:blanco1}. The outliers in the FGK range with shorter periods are likely to be close binaries that have been spun up by tidal interactions 
\citep{zahn-1977:tides-close-binaries, simonian-2019:binary-spinup}. Later type M-dwarfs, on the other hand, present a wide range of rotation periods.
We find that K2-136, which is a K-dwarf with a period of $13.6^{+2.2}_{-1.5}$ days, fits well within this tight relation, and thus has a rotation period that is typical for a Hyades member of this spectral type. The model by \citet{johnstone-2021} is also in good agreement with the star's rotation period.

In Fig.\,\ref{fig:hyades} (right hand panel) we also plot X-ray activity $L_\text{X}/L_\text{bol}$ against BP-RP colour for these Hyades stars. The stars we identified as spun-up binaries are generally more X-ray active than other FGK stars, which is expected as faster rotators are more X-ray active \citep{walter-1978:binary-spinup-xrays, dempsey-1993:binary-spinup-xrays, dempsey-1997:binary-spinup-xrays}.
Our measured X-ray activity of K2-136 from Sect.\,\ref{sec:xray-spectrum} is also plotted on Fig.\,\ref{fig:hyades}, as is the predicted X-ray activity from the \citet{johnstone-2021} model.
It can be seen that our measured X-ray activity of $1.77\times10^{-5}$ is a much better match to the X-ray detections and upper limits of similar Hyades K-dwarfs (with detections around $\sim2\times10^{-5}$) than it is to the model prediction, which lies between the measured single stars and close binaries at $6.6 \times 10^{-5}$.

For this reason, we favour the analysis of photoevaporation we carried out in Sect.\,\ref{sec:alternative}, where we scaled the \citet{johnstone-2021} model to our observed X-ray luminosity. 

More generally, Fig.\,\ref{fig:hyades} shows that the rotation-activity relation employed by \citet{johnstone-2021} is systematically over-predicting the X-ray luminosities of single G and K stars in the Hyades. This may be due to contamination by close binaries of the sample of stars used to determine the rotation-activity relation.

\subsection{Evaporation history of the three planets}
\label{sec:evaporation}

In Fig.\,\ref{fig:valley-evo} we present a period-radius plot with the locations of the three planets as observed at the present time, together with their evaporative pasts and futures as modelled in Sect.\,\ref{sec:results}.

\subsubsection{K2-136\,b}

We find that K2-136\,b is most consistent with an Earth-sized rocky world: an apparently typical member of the large population of small dense planets below the radius-period valley.
Our simulations show that its primordial envelope, if any, was lost shortly after the dispersal of the protoplanetary disc. 
Assuming this primordial envelope made up 1\% of its mass (mirroring its sibling K2-136\,c) we find that K2-136\,b would have formed with a radius of $2.8\,\Rearth$ adopting the envelope model by \citet{owen-wu-2017}, making it a sub-Neptune above the radius valley (Fig.\,\ref{fig:valley-evo}).

\subsubsection{K2-136\,c}

In contrast, K2-136\,c remains a sub-Neptune above the radius valley, with a thick envelope making up of 1--2\% of its mass. All our evaporation models suggest that this planet is stable against evaporation.
In large part, this is due to its relatively high mass of $18\,\Mearth$ measured with HARPS-N (Sect.\,\ref{sec:planets}).

Interestingly, this high mass implies a core radius of $2.4\,\Rearth$ (see Sect.\,\ref{sec:planets}), which would place K2-136\,c above the radius valley even if its envelope were entirely stripped (Fig.\,\ref{fig:valley-evo}). In such a case, without a mass measurement, the planet might be misidentified as a lower-mass planet with a thick envelope, leading to erroneous conclusions about the efficiency of envelope evaporation. This underlines the importance of mass measurements. It also illustrates that a location above the radius valley does not in itself require a planet to have a gaseous envelope. 

\begin{figure}
    \includegraphics[width=\columnwidth]{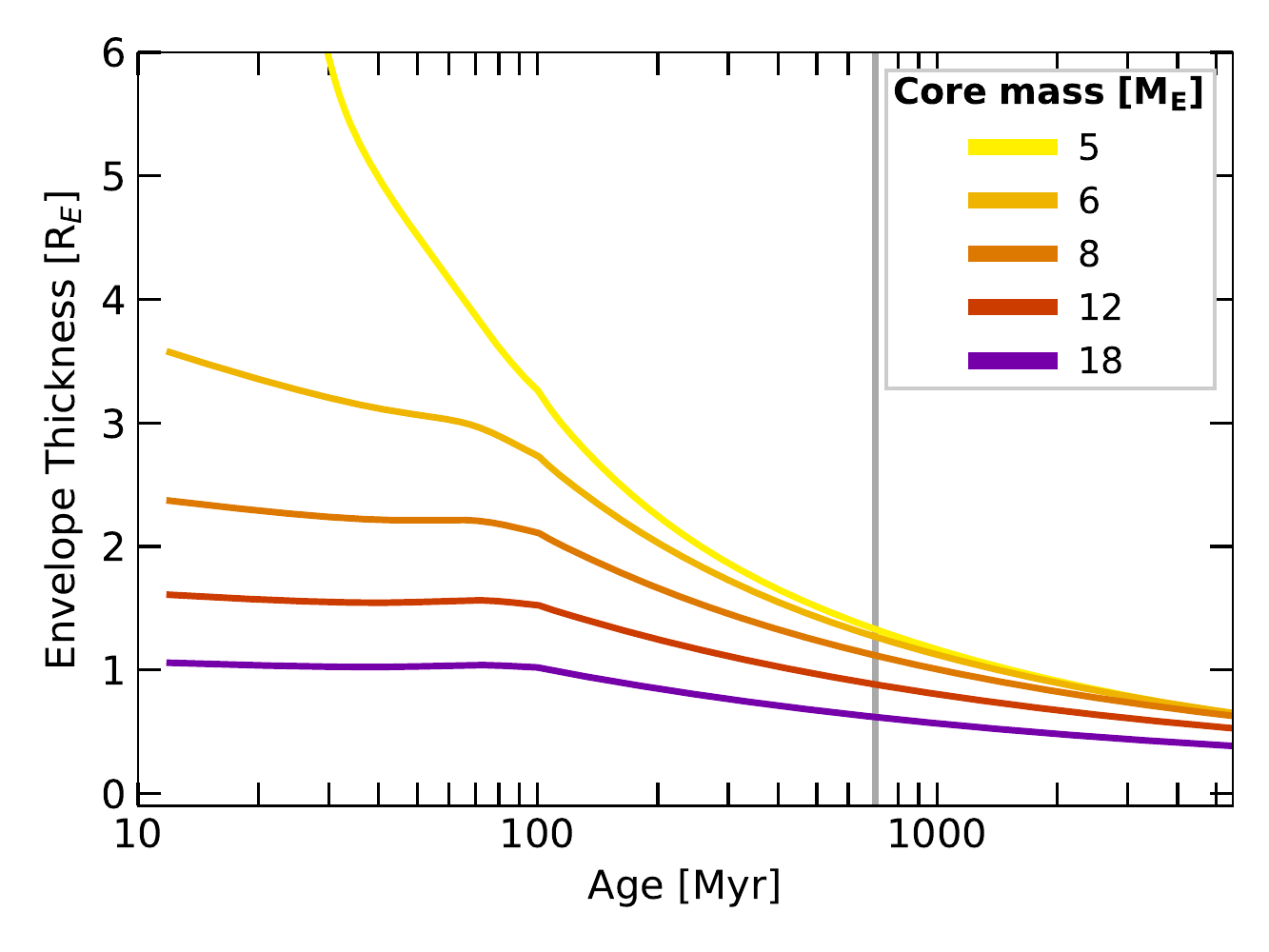}
    \caption{Plot of planet radius against age showing evolution of K2-136\,c with a radius of $3\,\Rearth$ and a range of planet masses between $5$ and $18\,\Mearth$, motivated by the variety of masses observed on exoplanet populations for planets of this size.}
    \label{fig:c-mcores}
\end{figure}

Unfortunately, we do not have mass measurements for most Kepler planets, 
and it is informative to consider how we would have interpreted K2-136\,c in the absence of a mass measurement. 
In that case we would have turned to a mass-radius relation such as that by \citet{otegi-2020}, which would point to a mass between 5 and 20\,$\Mearth$.
As an experiment, we explore the photoevaporation history of K2-136\,c as a function of its core mass, as we did for K2-136\,d in Sect.\,\ref{sec:histories-d}. The results of such an analysis using the \citet{kubyshkina-2018} model are plotted in Fig.\,\ref{fig:c-mcores}, which shows that the low end of the mass range implies unrealistically rapid evaporation at early times 
(see Sect.\,\ref{sec:dis-d}).
Photoevaporation models are therefore capable of placing a lower limit on the mass of a planet that is tighter than can be determined from a mass-radius relation alone. Similarly, our photoevaporation model would also rule out a scenario in which the planet was already a stripped core, since the implied higher core mass is shown to be stable to atmospheric escape.

\subsubsection{K2-136\,d}
\label{sec:dis-d}

As seen in Fig.\,\ref{fig:valley-evo}, K2-136\,d currently lies just below the valley, and its uncertain mass leads to a variety of possible internal structures, ranging from a $1.3\,\Mearth$ planet with a tenuous atmosphere (motivated by the mass upper limits; Sect.\,\ref{sec:planets}) to a completely rocky world with a mass of $3.6\,\Mearth$ (which is in tension with the HARPS-N upper limit).

According to our favoured evolution scenario, where we scale the XUV track to match our X-ray observations (Sects.\,\ref{sec:alternative}\,\&\,\ref{sec:hyades}), photoevaporation suggests the planet mass must be at least $2.0-2.5\,\Mearth$ in order to maintain a gaseous envelope to the present day (Fig.\,\ref{fig:evo-d-mcores-lowxuv}).   
Backwards evolution of these models suggest an initial state between a super-Earth with envelope mass fraction $\sim0.2\%$ in the middle of radius valley, to a sub-Neptune with a sizeable envelope of $1\%$ mass fraction, more akin to K2-136\,c.

For the lower-mass models, evolution backwards in time requires rapidly growing envelopes, with the mass of the envelope quickly becoming greater than the core mass (e.g. Fig.\,\ref{fig:evo-d-mcores-lowxuv}).
We have deemed these scenarios to be unphysical because in forward evolution they require unrealistic fine-tuning to their starting parameters to match the planet at the present time --- with a precise starting envelope only marginally lighter than the self-gravitating envelopes on gas giants, which are stable against evaporation \citep{murray-clay-2009:hot-jupiters-stable}.

\subsubsection{Alternative composition}

All of our models in Sects.\,\ref{sec:sim}\,\&\,\ref{sec:results} assume a planet with a rocky core, with Earth-like composition, and a gaseous envelope dominated by hydrogen and helium.
However, the low density of K2-136\,d implied by the 1-$\sigma$ mass upper limit of $1.3\,\Mearth$ (Sect.\,\ref{sec:planets}) could also be explained by a composition that included a significant proportion of water \citep[e.g.][]{zeng-2019:water-valley, venturini-2020:water-valley}.
Indeed, the mass-radius relations by \citet{owen-wu-2013}, which allow for varying water, iron, and silicate mass fractions within the core, show that a composition of 50\% ice and 50\% rock would reproduce a mass of $1.3\,\Mearth$ without any gaseous envelope. 
This scenario would also be compatible with a primordial gaseous envelope that has already been completely removed.
Of course, a significant water content is also possible for K2-136\,c.

The presence of considerable water compositions in close-in exoplanets, though, is contested.
\citet{rogers-owen-2021} produced an ensemble of synthetic planets using distributions of core masses, core compositions, and envelope mass fractions fitted to the observed CKS sample of planets, and evolved them under photoevaporation. They found that the Kepler planets are most consistent as rock-iron cores that are either bare or surrounded by gaseous envelopes. Their results also indicate little evidence of the widespread presence of water-rich worlds, as these planets would produce a radius-valley at greater radii than observed under photoevaporation, or no valley at all in the absence of mass loss.

\subsubsection{Multi-planet systems}

The presence of the Earth-sized K2-136\,b and the sub-Neptune K2-136\,c alone 
would be consistent with the picture that the architectures of multi-planet systems tend towards size ordering, with inner planets below the valley that are stripped, and less-irradiated outer planets above the valley that maintain a gaseous envelope \citep{ciardi-2013:architecture, millholland-2017:architecture, weiss-2022:architecture}. The additional presence of the super-Earth K2-136\,d below the valley as the outermost planet is intriguing.

\begin{figure}
    \includegraphics[width=\columnwidth]{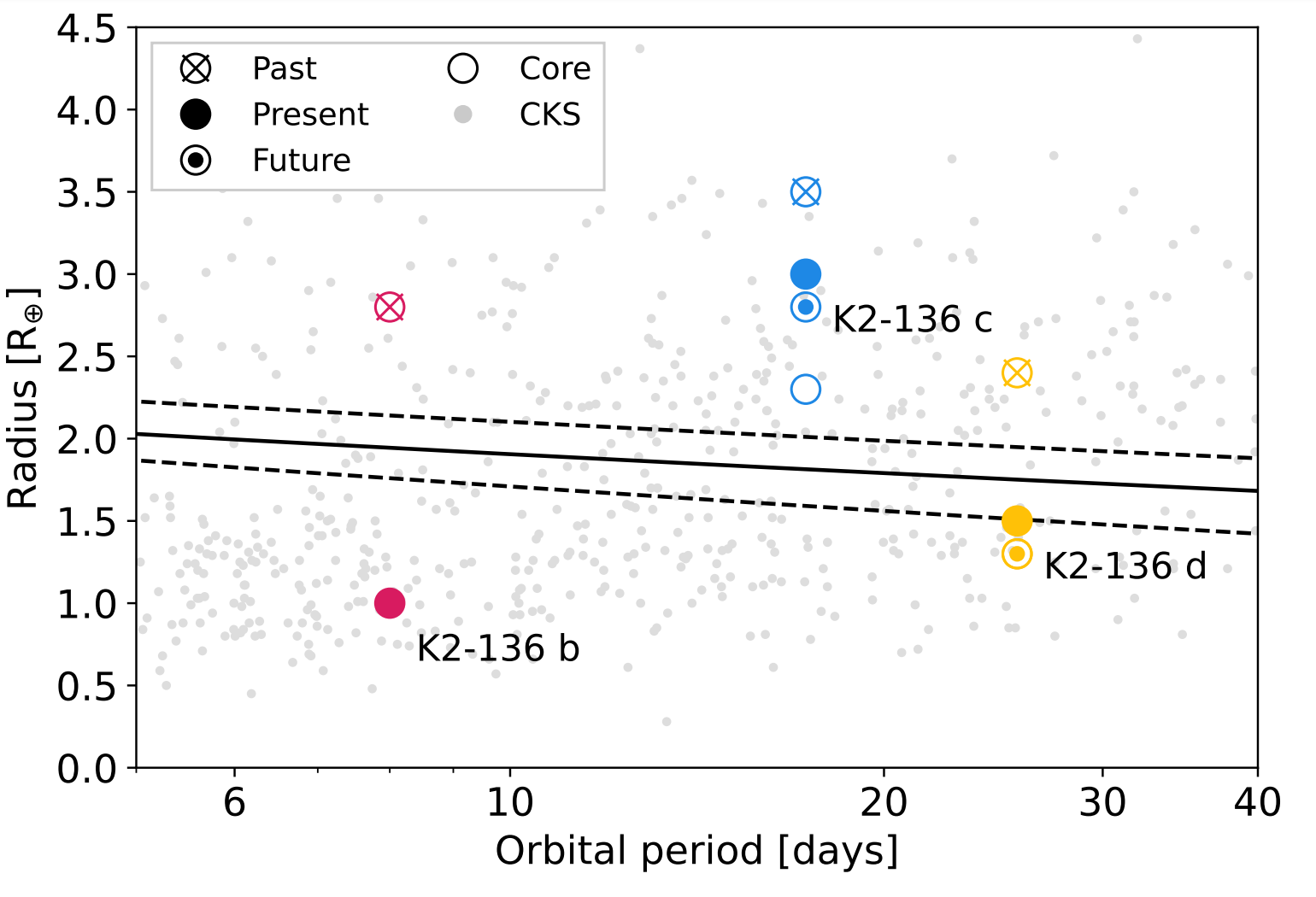}
    \caption{Plot of orbital period against planetary radius.
    The radii of the K2-136 planets at different points throughout their history are plotted as circles in red (K2-136\,b), blue (K2-136\,c), and yellow (K2-136\,d). The circles with a cross, a solid colour, and a dot represent the past, present, and future states, respectively. The circles with no filling represent the planets' core alone. The initial size of K2-136\,d was calculated using a core mass of 2.5 $\Mearth$, following our results in Sect.\,\ref{sec:alternative}. The past radius of K2-136\,b was estimated from an envelope mass fraction of 1.5\%, mirroring K2-136\,c and d.
    The California-\textit{Kepler} Survey (CKS) planets \citep{fulton-2017} are plotted as grey points, and the radius-period valley as defined by \citet{van-eylen-2018} is marked with a black line (with the 1$\sigma$ errors as the dashed lines).}
    \label{fig:valley-evo}
\end{figure}

Based on our modelling, we have plotted the likely past, present and future architectures of the K2-136 planetary system in Fig.\,\ref{fig:valley-evo}. 
Overall, we find that all three planets are consistent with starting out as sub-Neptunes above the radius valley with envelope mass fractions of $\sim$1\%. We find that planet d is now below the radius valley because it lost most of its envelope due to a low core mass. This makes it more susceptible to photoevaporation than planet c, despite experiencing weaker XUV irradiation.  

Data from the NASA Exoplanet Archive
\footnote{The NASA Exoplanet Archive can be accessed on \url{https://exoplanetarchive.ipac.caltech.edu/}}
reveals that about 30 multi-planet systems ($\sim3\%$ of those discovered to date) host a planet above the valley ($R_\text{p} \ge 2.0\,\Rearth$) interior to a another planet below the valley ($R_\text{p} \le 1.6\,\Rearth$), 13 of which are three-planet systems, akin to K2-136. This suggests that the architecture of the K2-136 system is relatively rare, but by no means unique.
In such systems we need to understand how the outer planet formed with a lower mass core, when both planet radii and masses tend to increase with separation within the protoplanetary disc \citep{weiss-2018:cks-peas, millholland-2017:architecture}.

Our results also illustrate how photoevaporation is able to produce a diverse set of architectures, since it is sensitive both core masses and XUV irradiation history. Our results are also sensitive to the choice of atmospheric escape model, suggesting that observations of multi-planet systems can distinguish between these models. We find that the choice of envelope structure model is less important as the models are largely consistent with one another in both the higher mass regime, as seen with K2-136\,c in Fig.\,\ref{fig:evo-c}, and the low mass regime, with K2-136\,d in Fig.\,\ref{fig:evo-d-light}.

As pointed out by \citet{campos-estrada-20}, multi-planet systems provide the most sensitive tests of photoevaporation-driven evolution, especially when the planets straddle the radius-period valley. Studies of multi-planet systems in open clusters, such as ours, have the added advantages of known  ages and a set of sibling stars to assess the average XUV activity of the host star. Future studies of additional systems at a range of ages, including very young planets \citep[e.g.][]{poppenhaeger-2021:multiplanet-evolution}, and planets with measured masses, have the potential to distinguish between evaporation models and directly determine the efficiency of the mass loss.

\section{Conclusion}\label{sec:conclusion}

We have investigated the past and future evolution of the planetary system around K2-136, which is a Hyades cluster member hosting three transiting exoplanets spanning the radius-period valley. The system is intriguing because the innermost and outermost planets lie below the radius valley, whereas the middle planet lies above the valley. Its presence in the Hyades open cluster provides both a known age of 700\,Myr and a set of similar stars with which to compare its XUV emission. 

We employed an \xmm\ observation to measure the X-ray luminosity K2-136, which we find to be typical of similar single stars in the Hyades cluster. This X-ray luminosity is a factor of 2.7 lower than predicted by the model of \citet{johnstone-2021}, suggesting that existing rotation-activity relations may be biased by close binaries. 

Using both the measured (Sect.\,\ref{sec:xray-spectrum}) and predicted (Sect.\,\ref{sec:predicted-xuv}) X-ray activity of the star, we modeled the photoevaporative evolution of all three planets forwards and backwards in time from their present age. 

We employed a range of planet structure models, and a range of mass-loss prescriptions, including core-powered mass loss. Overall, we found that all three planets are consistent with having formed as sub-Neptunes above the radius-period valley, with envelope mass fractions of around 1\%.

We find that the innermost planet, K2-136\,b, will have lost any primordial gaseous envelope early in its evolution, regardless of choice of structure model or mass-loss prescription. 
In contrast, the relatively high measured mass of the middle planet, K2-136\,c, results in an envelope that is stable to mass loss, again regardless of structure model or mass-loss prescription. 

The outer planet, however, K2-136\,d, has a history and future that is sensitive to the XUV irradiation and the choice of mass-loss prescription. 
Using our measured X-ray luminosity to scale the XUV irradiation history, we find that some core-powered and energy-limited photoevaporative mass loss prescriptions are both consistent with formation as a sub-Neptune with a core mass that conforms to upper limits from HARPS-N observations.
The models of \citet{kubyshkina-2018} and \citet{salz-2016}, however, are only consistent with core masses at the upper end of the allowed range from HARPS-N,
greater than $2.5\,\Mearth$,
and favour a fully-stripped massive core at the present time (which is in tension with the HARPS-N data). A more precise mass measurement (or limit) for this planet would tightly constrain the allowed sets of photoevaporation models. 

Our work has shown that studies of multi-planet systems spanning the radius-period valley can provide meaningful constraints on the physical processes controlling mass loss from planetary envelopes. 
This is especially true where masses can be measured for the planets and where the age of the system is tightly constrained.
As we have shown, membership of an open cluster is also valuable in testing whether the X-ray emission of a host star is typical for its age and mass --- marginalising over variability of the X-ray emission.

\section*{Acknowledgements}

The work presented here was supported by UK Science and Technology Facilities (STFC).
JFF acknowledges support from an STFC studentship under grant ST/W507908/1, and PJW acknowledges support under STFC consolidated grant ST/T000406/1.

This research was based on observations obtained with XMM-Newton, an ESA science mission with instruments and contributions directly funded by ESA Member States and NASA.

This research made use of the Python packages \texttt{\,numpy} \citep{numpy}, \texttt{\,astropy} \citep{astropy}, \texttt{\,scipy} \citep{scipy}, \texttt{\,matplotlib} \citep{matplotlib}, and \texttt{\,Mors} \citep{johnstone-2021}.

This research has made use of the NASA Exoplanet Archive, which is operated by the California Institute of Technology, under contract with the National Aeronautics and Space Administration under the Exoplanet Exploration Program.

\section*{Data Availability}

The \xmm data used in this work is publicly available at the \xmm Science Archive (XSA) \footnote{\url{https://www.cosmos.esa.int/web/xmm-newton/xsa}}, under the observation ID 0824850201 (target: LP 358--348, PI: Wheatley). 




\bibliographystyle{mnras}
\bibliography{biblio} 




\bsp	
\label{lastpage}
\end{document}